\documentclass[a4paper, 11pt]{article}
\usepackage{indentfirst}
\usepackage{graphicx}
\usepackage{subfigure}
\usepackage{caption2}
\usepackage{overpic}
\usepackage{psfrag}
\usepackage[dvips]{hyperref}
\usepackage{amsmath, amssymb, amsthm}

\newcommand\bbR{\mathbb{R}}
\newcommand\bbN{\mathbb{N}}
\newcommand\bxi{\boldsymbol{\xi}}
\newcommand\bx{\boldsymbol{x}}
\newcommand\bv{\boldsymbol{v}}
\newcommand\bu{\boldsymbol{u}}
\newcommand\bw{\boldsymbol{w}}

\newcommand\bF{\boldsymbol{F}}

\newcommand\bH{\boldsymbol{H}}

\newcommand\dd{\,\mathrm{d}}
\newcommand\He{\mathit{He}}
\newcommand\Kn{\mathit{Kn}}
\newcommand\Ma{\mathit{Ma}}

\newcommand\NRxx{NR$xx$}
\newcommand\bNRxx{\texorpdfstring{{\bf NR}$\boldsymbol{xx}$}{NR$xx$}}

\numberwithin{equation}{section}

\graphicspath{{images/}}

{\theoremstyle{remark} \newtheorem{remark}{Remark}}

\title{The {\NRxx} Method for Polyatomic Gases}

\author{Zhenning Cai\thanks{School of Mathematical Sciences, Peking
    University, Beijing, China, email: {\tt caizn@pku.edu.cn}.},~~
  Ruo Li\thanks{CAPT, LMAM \& School of Mathematical Sciences, Peking
    University, Beijing, China, email: {\tt rli@math.pku.edu.cn}.}}

\begin{document}
\maketitle

\begin{abstract}
  In this paper, we propose a numerical regularized moment method to
  solve the Boltzmann equation with ES-BGK collision term to simulate
  polyatomic gas flows. This method is an extension to the polyatomic
  case of the method proposed in \cite{NRxx_new}, which is abbreviated
  as the {\NRxx} method in \cite{Cai}. Based on the form of the
  Maxwellian, the Laguerre polynomials of the internal energy
  parameter are used in the series expansion of the distribution
  function. We develop for polyatomic gases all the essential
  techniques needed in the {\NRxx} method, including the efficient
  projection algorithm used in the numerical flux calculation, the
  regularization based on the Maxwellian iteration and the order of
  magnitude method, and the linearization of the regularization term
  for convenient numerical implementation. Meanwhile, the particular
  integrator in time for the ES-BGK collision term is put forward. The
  shock tube simulations with Knudsen numbers from $0.05$ up to $5$
  are presented to demonstrate the validity of our method.  Moreover,
  the nitrogen shock structure problem is included in our numerical
  experiments for Mach numbers from $1.53$ to $6.1$.

\vspace*{4mm}
\end{abstract}

\section{Introduction}
The kinetic theory has long been playing an important role in the
rarefied gas dynamics. As a mesocopic theory standing between the
fluid dynamics and the molecular dynamics, the kinetic theory is built
on the basis of the Boltzmann equation, which uses a distribution
function to give a statistical description of the distribution of
microscopic particles' velocities. In 1940s, Grad \cite{Grad} proposed
the idea using the Hermite expansion to approximate the
distribution function, and a 13-moment theory was given in detail in
\cite{Grad}. Recently, based on the idea of Grad, systems with large
numbers of moments together with their numerical schemes are
considered in \cite{NRxx, NRxx_new, Cai}, where some regularizations
inspired by \cite{Struchtrup2003, Struchtrup2004} are also taken into
account. In \cite{Cai}, the numerical regularized moment method is
abbreviated as the {\NRxx} method. However, all these works
concentrate only on the monatomic gases, and in this paper, we will
develop the {\NRxx} method for the polyatomic case.

The study to apply the moment method to polyatomic gases can be traced
back to McCormack \cite{McCormack}, where a 17-moment model was
proposed. As far as we know, the most recent polyatomic extension of
Grad's 13-moment equations is the work of Mallinger \cite{Mallinger},
whose system contains only 14 moments. In both models, a great amount
of work is devoted to the deduction of the collision terms. In order
to generalize the moment theory to large number of moments, we prefer
a BGK-like simplified collision operator. As in the monatomic case,
the simplest BGK model fails to give correct heat conduction, and for
polyatomic gases, it also gives incorrect relaxation collision number,
resulting in qualitative errors in temperatures compared with the
Boltzmann equation \cite{Bourgat}. Possible alternatives include the
Rykov model \cite{Rykov} and the ES-BGK model \cite{Andries}, which
incorporate physical Prandtl number and relaxation collision number
into the collision term. In this work, our investigation is restricted
to the ES-BGK model.

For polyatomic gases, besides the velocities of microscopic particles,
an additional nonnegative ordinate representing the energy of internal
degrees of freedom appears in the distribution function. Thus, in
order to expand the distribution function into series, the basis
functions are chosen as a combination of Hermite polynomials and
Laguerre polynomials with proper translation and scaling based on the
macroscopic velocity and translational and rotational temperatures of
the gas. By considering the coefficients of the basis functions as
moments, a system with infinite number of moment equations is
derived. A moment closure is then followed to truncate the system with
infinite equations and get a system with only finite equations. The
framework for the moment closure is the same as \cite{NRxx_new}:
\begin{enumerate}
\setlength\itemsep{0cm}
\item the Maxwellian iteration is applied to determine the order of
  magnitude for each moment;
\item by dropping higher order terms, the truncated moments are
  expressed by moments with lower orders;
\item for easier numerical implementation, the expression is
  linearized around a Maxwellian.
\end{enumerate}
However, the details of the Maxwellian iteration are significantly
different. In the polyatomic case, the iteration is much more
complicated than the monatomic case because of the existence of both
translational and rotational temperatures in the basis functions, and
the process should be conducted carefully. Moreover, for the ES-BGK
model, analysis on the moments of the Gauss distribution also
increases the complexity. Fortunately, the final result remains a
similar form as simple as in \cite{NRxx_new}.

As to the numerical scheme, the general framework in \cite{Cai} is
applicable. A split scheme is applied to divide the transportation
part and the collision part, and the transportation part is processed
by a finite volume method. Recalling that a special ``projection''
introduced in \cite{NRxx, Cai} is required in the calculation of
numerical fluxes, we further develop this technique to the polyatomic
case in this paper. Meanwhile, the polyatomic ES-BGK collision term
can no longer be solved analytically as the BGK operator \cite{NRxx},
and the Crank-Nicolson scheme is applied to ensure the unconditional
numerical stability. Our numerical experiments show that our scheme
correctly converges to the solution of the Boltzmann equation as the
number of moments increases. The distinction between BGK and ES-BGK
models, together with the relation between monatomic and polyatomic
cases, is illustrated by the numerical results of shock tube
problems. Also, we apply the {\NRxx} method to the nitrogen shock
structure problem, and the results are comparable to the experimental
data.

The rest of this paper is arranged as follows: in Section
\ref{sec:ES-BGK}, a brief review of the polyatomic ES-BGK model is
given. In Section \ref{sec:NRxx}, the polyatomic {\NRxx} method is
introduced comprehensively, and in Section \ref{sec:num}, a number of
numerical experiments are carried out to validate our algorithm.  As a
summation, some concluding remarks are given in Section
\ref{sec:conclusion}. Finally, some involved calculations are
collected in the appendix for better readability to the body matter.


\section{The ES-BGK Boltzmann equation for polyatomic gases} \label{sec:ES-BGK}
The ES-BGK model for polyatomic gases, which gives correct
Navier-Stokes heat conduction compared with the BGK model, has been
deduced in \cite{Andries, Brull}. The polyatomic ES-BGK Boltzmann
equation reads
\begin{equation} \label{eq:ES-BGK}
\frac{\partial f}{\partial t} + \bxi \cdot \nabla_{\bx} f =
  \mathrm{Pr} \cdot \frac{p}{\mu}(G - f),
\end{equation}
where $f$ denotes the molecule distribution, which is a positive
function $f = f(t,\bx,\bxi,I)$ with $\bx,\bxi \in \bbR^3$ and $t,I \in
\bbR^+$. The parameters $t$, $\bx$ and $\bxi$ stand for the time,
spatial position and microscopic molecule velocity respectively, and
$I$ is an internal energy parameter. In the right hand side of
\eqref{eq:ES-BGK}, $\mathrm{Pr}$ is the Prandtl number, $p$ is the
pressure, and $\mu$ denotes the viscosity coefficient. $G$ is a
generalized Gaussian defined as
\begin{equation} \label{eq:G}
G(t,\bx,\bxi,I) =
  \frac{\rho \Lambda_{\delta}}
    {\sqrt{\det(2\pi \mathcal{T})}(RT_{\mathrm{rel}})^{\delta/2}}
  \exp \left(
    -\frac{1}{2} (\bxi - \bu)^T \mathcal{T}^{-1} (\bxi - \bu)
    - \frac{I^{2/\delta}}{RT_{\mathrm{rel}}}
  \right).
\end{equation}
Here $\delta$ is the total number of molecular internal degrees of
freedom, and $R$ is the gas constant. The density $\rho$ and the
macroscopic velocity $\bu$ are related to the distribution function
$f$ through
\begin{equation} \label{eq:rho_u}
\rho = \int_{\bbR^3 \times \bbR^+} f \dd \bxi \dd I, \qquad
\bu =
  \frac{1}{\rho} \int_{\bbR^3 \times \bbR^+} \bxi f \dd \bxi \dd I,
\end{equation}
and $T_\mathrm{rel}$ is a relaxation temperature
\begin{equation}
T_\mathrm{rel} = Z^{-1} T_{\mathrm{eq}} + (1-Z^{-1}) T_{\mathrm{int}},
\end{equation}
where $Z$ is the relaxation collision number. For polyatomic gases,
three temperatures are used frequently, including the translational
temperature $T_{\mathrm{tr}}$, the internal temperature
$T_{\mathrm{int}}$, and the equilibrium temperature $T_{\mathrm{eq}}$.
They are defined by
\begin{align}
\label{eq:T_tr}
T_{\mathrm{tr}} & = \frac{1}{3 \rho R} \int_{\bbR^3 \times \bbR^+}
  |\bxi - \bu|^2 f \dd \bxi \dd I, \\
\label{eq:T_int}
T_{\mathrm{int}} & = \frac{2}{\delta \rho R} \int_{\bbR^3 \times \bbR^+}
  I^{2/\delta} f \dd \bxi \dd I, \\
\label{eq:T_eq}
T_{\mathrm{eq}} & =
  (3 T_{\mathrm{tr}} + \delta T_{\mathrm{int}}) / (3 + \delta).
\end{align}
And the pressure $p$ is obtained from the ideal gas law:
\begin{equation} \label{eq:pressure}
p = \rho R T_{\mathrm{eq}}.
\end{equation}
Now it only remains to define $\Lambda_{\delta}$ and $\mathcal{T}$:
\begin{equation}
\Lambda_{\delta} = \left[
  \int_{\bbR^+} \mathrm{e}^{-I^{2/\delta}} \dd I
\right]^{-1}, \quad
\mathcal{T} = (1 - Z^{-1})
  [(1-\nu) R T_{\mathrm{tr}} \mathrm{Id} + \nu \Theta / \rho]
  + Z^{-1} R T_{\mathrm{eq}} \mathrm{Id},
\end{equation}
where
\begin{equation} \label{eq:Theta_nu}
\Theta = \int_{\bbR^3 \times \bbR^+}
  (\bxi - \bu) \otimes (\bxi - \bu) f \dd \bxi \dd I, \quad
\nu = \frac{1 - \mathrm{Pr}^{-1}}{1 - Z^{-1}},
\end{equation}
and $\mathrm{Id}$ stands for the identity matrix.

\section{The {\bNRxx} method for polyatomic ES-BGK equation} \label{sec:NRxx}
In this section, we are going to extend the {\NRxx} method proposed in
\cite{NRxx, NRxx_new} to the polyatomic case, which includes the
following steps:
\begin{enumerate}
\setlength\itemsep{0cm}
\item The distribution function is expanded into a series with
  specially selected basis functions.
\item A system with infinite number of moment equations is deduced.
\item The moment system is truncated at a certain place and made closed
  by regularization.
\item The regularization term is linearized in order to simplify the
  numerical implementation.
\item The numerical method is carried out following \cite{Cai}.
\end{enumerate}
The details are introduced in the following five subsections.

\subsection{Spectral representation of the velocity space} \label{sec:spectral}

In the {\NRxx} method for the monatomic gases, the Hermite polynomials
have been employed to construct the basis functions of the velocity
space, since Hermite polynomials are orthogonal over the region
$(-\infty, +\infty)$. For the polyatomic distribution function, since
$I \in \bbR^+$, we use the Laguerre polynomials, which are orthogonal
over the region $[0, +\infty)$, as the basis functions in the
ordinate $I$. Thus the basis function has the following form:
\begin{equation} \label{eq:bas_fun}
\begin{split}
\psi_{\alpha, k, T_{\mathrm{tr}}, T_{\mathrm{int}}} (\bv, J) & =
  \frac{2}{\delta} \left( \gamma_k^{(m)} \right)^{-1}
  (R T_{\mathrm{int}})^{-(\delta/2+k)} L_k^{(m)}(J) \exp(-J) \cdot{} \\
& \qquad \left( \sqrt{2 \pi} \right)^{-3}
  (R T_{\mathrm{tr}})^{-\frac{|\alpha|+3}{2}}
  \prod_{d=1}^3 \He_{\alpha_d}(v_d) \exp \left(-\frac{v_d^2}{2}\right),
\end{split}
\end{equation}
where $\alpha = (\alpha_1, \alpha_2, \alpha_3)$ is a multi-index, and
\begin{gather}
\label{eq:m_gamma}
m = \delta / 2 - 1, \qquad
  \gamma_k^{(m)} = \frac{\Gamma(m + k + 1)}{\Gamma(k + 1)}, \\
\label{eq:Laguerre}
L_k^{(m)}(J) = \frac{J^{-m}\mathrm{e}^J}{k!}
  \frac{\mathrm{d}^k}{\mathrm{d} J^k} (\mathrm{e}^{-J} J^{k+m}), \\
\label{eq:Hermite}
\He_n(x) = (-1)^n \exp \left( \frac{x^2}{2} \right)
  \frac{\mathrm{d}^n}{\mathrm{d} x^n}
  \exp \left( -\frac{x^2}{2} \right).
\end{gather}
Some properties of the Laguerre polynomials $L_k^{(m)}$ and the
Hermite polynomials $\He_n$ can be found in Appendix \ref{sec:orth}.
With equation \eqref{eq:bas_fun}, the distribution function $f(\bxi,
I)$ is expanded as
\begin{equation} \label{eq:expansion}
f(\bxi, I) = \sum_{\alpha \in \bbN^3} \sum_{k \in \bbN}
  f_{\alpha,k} \psi_{\alpha,k,T_{\mathrm{tr}}, T_{\mathrm{int}}}
  \left(
    \frac{\bxi - \bu}{\sqrt{R T_{\mathrm{tr}}}},
    \frac{I^{2/\delta}}{R T_{\mathrm{int}}}
  \right).
\end{equation}
Let us consider the general case when $T_{\mathrm{tr}}$,
$T_{\mathrm{int}}$ and $\bu$ have no relation with the distribution
function $f$. Using the orthogonality of the Laguerre and Hermite
polynomials, we can deduce that
\begin{subequations} \label{eq:mnts}
\begin{gather}
\int_{\bbR^3 \times \bbR^+} f \dd \bxi \dd I = f_{0,0}, \\
\int_{\bbR^3 \times \bbR^+} \xi_j f \dd \bxi \dd I
  = f_{0,0} u_j + f_{e_j,0}, \quad j = 1,2,3, \\
\int_{\bbR^3 \times \bbR^+} I^{2/\delta} f \dd \bxi \dd I
  = \frac{1}{2} \delta R T_{\mathrm{int}} f_{0,0} - f_{0,1}, \\
\int_{\bbR^3 \times \bbR^+} \frac{1}{2} |\bxi|^2 f \dd \bxi \dd I
  = \frac{1}{2} f_{0,0} |\bu|^2 + \sum_{j=1}^3 \left(
    \frac{1}{2} R T_{\mathrm{tr}} f_{0,0} +
    u_j f_{e_j,0} + f_{2e_j,0}
  \right).
\end{gather}
\end{subequations}
If $\bu$ is the macroscopic velocity and $T_{\mathrm{tr}}$,
$T_{\mathrm{int}}$ are the translational and internal temperatures for
the distribution $f$, using \eqref{eq:rho_u}, \eqref{eq:T_tr}, 
\eqref{eq:T_int} and \eqref{eq:mnts}, we conclude
\begin{equation} \label{eq:normal_rep}
f_{0,0} = \rho, \qquad
  f_{e_j,0} = f_{0,1} = \sum_{d=1}^3 f_{2e_d,0} = 0,
  \quad j = 1,2,3.
\end{equation}
If \eqref{eq:normal_rep} is satisfied, then \eqref{eq:expansion} is
called as a \emph{normal representation} of $f$. If
\eqref{eq:expansion} is not a normal representation, then the density,
momentum and translational and internal energies can be easily
calculated through \eqref{eq:mnts}. For a normal representation, we
have
\begin{equation} \label{eq:Theta}
\Theta_{ij} = (1 + \delta_{ij}) f_{e_i + e_j,0}
  + \delta_{ij} \rho R T_{\mathrm{tr}}, \qquad i,j=1,2,3.
\end{equation}
where $\Theta$ is defined in \eqref{eq:Theta_nu}.

\subsection{The moment equations for the ES-BGK model}
In this section, we are going to derive equations for the moment set
$\{f_{\alpha,k}\}$. The general strategy is to substitute
\eqref{eq:expansion} into \eqref{eq:ES-BGK}, and then match the
coefficients of the same basis functions. For the left hand side of
\eqref{eq:expansion}, the process is similar as that in
\cite{NRxx_new}, and the detailed derivation can be found in Appendix
\ref{sec:mnt_eqs}. Suppose $G$ has the following expansion:
\begin{equation} \label{eq:G_expand}
G(t, \bx, \bxi, I) = \sum_{\alpha \in \bbN^3} \sum_{k \in \bbN}
  G_{\alpha,k}(t,\bx) \psi_{\alpha,k,T_{\mathrm{tr}},T_{\mathrm{int}}}
  \left(
    \frac{\bxi - \bu}{\sqrt{R T_{\mathrm{tr}}}},
    \frac{I^{2/\delta}}{R T_{\mathrm{int}}}
  \right),
\end{equation}
Then the analytical expressions of the moment equations are obtained
as
\begin{equation} \label{eq:mnt_eqs}
\begin{split}
& \frac{\partial f_{\alpha,k}}{\partial t}
  + \sum_{d=1}^3 \frac{\partial u_d}{\partial t} f_{\alpha-e_d,k}
  + \frac{1}{2} \frac{\partial (R T_{\mathrm{tr}})}{\partial t}
    \sum_{d=1}^3 f_{\alpha-2e_d,k}
  - (m+k) \frac{\partial (R T_{\mathrm{int}})}{\partial t}
    f_{\alpha, k-1} \\
& \quad + \sum_{j=1}^3 \bigg[ \left(
  R T_{\mathrm{tr}} \frac{\partial f_{\alpha-e_j,k}}{\partial x_j}
  + u_j \frac{\partial f_{\alpha,k}}{\partial x_j}
  + (\alpha_j + 1) \frac{\partial f_{\alpha+e_j,k}}{\partial x_j}
\right) \\
& \quad \quad + \sum_{d=1}^3 \frac{\partial u_d}{\partial x_j}
  \left( R T_{\mathrm{tr}} f_{\alpha-e_d-e_j,k} + u_j f_{\alpha-e_d,k}
    + (\alpha_j + 1) f_{\alpha-e_d+e_j,k} \right) \\
& \quad \quad + \frac{1}{2}
  \frac{\partial (R T_{\mathrm{tr}})}{\partial x_j} \sum_{d=1}^3 \left(
    R T_{\mathrm{tr}} f_{\alpha-2e_d-e_j,k}
    + u_j f_{\alpha-2e_d,k}
    + (\alpha_j + 1) f_{\alpha-2e_d+e_j,k}
  \right) \\
& \quad \quad -(m+k)\frac{\partial (R T_{\mathrm{int}})}{\partial x_j}
  \left(
    R T_{\mathrm{tr}} f_{\alpha-e_j,k-1}
    + u_j f_{\alpha,k-1}
    + (\alpha_j + 1) f_{\alpha+e_j,k-1}
  \right) \bigg] \\
& \hspace{275pt} = \mathrm{Pr} \cdot \frac{p}{\mu}
  (G_{\alpha,k} - f_{\alpha,k}),
\end{split}
\end{equation}
where $f_{\beta,l}$ is taken as zero when $l$ or any of the components
of $\beta$ is negative, and $m$ is defined in \eqref{eq:m_gamma}.

Now we focus on the relation between $G_{\alpha,k}$ and
$f_{\alpha,k}$. The expression of $G$ \eqref{eq:G}---%
\eqref{eq:Theta_nu} and the equalities under normal representation
\eqref{eq:normal_rep} and \eqref{eq:Theta} show that $G_{\alpha,k}$
are functions of $\rho$, $T_{\mathrm{tr}}$, $T_{\mathrm{int}}$ and
$f_{e_i+e_j, 0}$ with $i,j=1,2,3$. Thus the system \eqref{eq:mnt_eqs}
is closed for $\alpha \in \bbN^3$ and $k = 0,1$, which means only the
expressions of $G_{\alpha,0}$ and $G_{\alpha,1}$ are needed. The
following results are trivial:
\begin{subequations} \label{eq:G_alpha_k}
\begin{gather}
\label{eq:G_0_0}
G_{0,0} = \int_{\bbR^3 \times \bbR^+} G \dd \bxi \dd I = \rho, \\
\label{eq:G_1_0}
G_{e_j,0} = \int_{\bbR^3 \times \bbR^+} \xi_j G \dd \bxi \dd I
  - G_{0,0} u_j = 0,\\
\label{eq:G_0_1}
\begin{split}
G_{0,1} &= \frac{\delta}{2} R T_{\mathrm{int}} G_{0,0} -
  \int_{\bbR^3 \times \bbR^+} I^{2/\delta} G \dd \bxi \dd I \\
&= \frac{\delta}{2} \rho
  [R T_{\mathrm{int}} - Z^{-1} R T_{\mathrm{eq}}
   - (1 - Z^{-1}) R T_{\mathrm{int}}] \\
&= \frac{\delta}{2Z} \rho (R T_{\mathrm{int}} - R T_{\mathrm{eq}}),
\end{split}
\end{gather}
\end{subequations}
where \eqref{eq:G_0_1} comes from the physical meaning of the
relaxation collision number. Moreover, since $G(\bxi, I)$ has the
form
\begin{equation}
G(\bxi,I) = G_1(\bxi) G_2(I),
\end{equation}
where
\begin{equation}
G_1(\bxi) = 
  \frac{\rho}{\sqrt{\det(2\pi \mathcal{T})}}
  \exp \left(
    -\frac{1}{2} (\bxi - \bu)^T \mathcal{T} (\bxi - \bu)
  \right), \quad
G_2(I) = 
  \frac{\Lambda_{\delta}}{(RT_{\mathrm{rel}})^{\delta/2}}
  \exp \left(
    - \frac{I^{2/\delta}}{RT_{\mathrm{rel}}}
  \right),
\end{equation}
and the basis functions $\psi_{\alpha, k, T_{\mathrm{tr}},
  T_{\mathrm{int}}}$ have the same structure (see \eqref{eq:psi})
\begin{equation}
\psi_{\alpha,k,T_{\mathrm{tr}}, T_{\mathrm{int}}}(\bv, J) =
  \psi_{1,\alpha,T_{\mathrm{tr}}}(\bv)
  \psi_{2,k,T_{\mathrm{int}}}(J),
\end{equation}
we can conclude
\begin{equation} \label{eq:G_alpha_1}
G_{\alpha,1} = C G_{\alpha,0},
\end{equation}
where $C$ is independent of $\alpha$. From \eqref{eq:G_0_0} and
\eqref{eq:G_0_1}, we find
\begin{equation} \label{eq:C}
C = \frac{\delta}{2Z} (R T_{\mathrm{int}} - R T_{\mathrm{eq}}).
\end{equation}
Until now, what remains is to work out the expressions of
$G_{\alpha,0}$ for $|\alpha| \geqslant 2$. This needs some involved
calculation with details in Appendix \ref{sec:G}. The final result is
in a recursive form as
\begin{equation} \label{eq:G_recur}
G_{\alpha,0} = \frac{1}{\alpha_i} \sum_{j=1}^3
  [(1-\mathrm{Pr}^{-1}) (\Theta_{ij} / \rho
    - R T_{\mathrm{tr}} \delta_{ij})
    + Z^{-1} (R T_{\mathrm{eq}} - R T_{\mathrm{tr}}) \delta_{ij}]
  G_{\alpha-e_i-e_j,0},
\end{equation}
where $i \in \{1,2,3\}$ such that $\alpha_i > 0$, and
$G_{\alpha-e_i-e_j,0}$ is taken as zero when $\alpha_j - \delta_{ij}
-1 < 0$. With \eqref{eq:G_1_0}, one can easily observe that
$G_{\alpha, 0} = 0$ when $|\alpha|$ is odd.

As the end of this subsection, we give the equations for the velocity
$\bu$ and the translational and internal temperatures
$T_{\mathrm{tr}}$, $T_{\mathrm{int}}$. The equation for $u_d$ can be
obtained by substituting $\alpha$ with $e_d$ in \eqref{eq:mnt_eqs},
and the result is
\begin{equation} \label{eq:u}
\rho \frac{\partial u_d}{\partial t} + \sum_{j=1}^3 \left(
  \rho u_j \frac{\partial u_d}{\partial x_j} +
  \frac{\partial \Theta_{j d}}{\partial x_j}
\right) = 0, \qquad j = 1,2,3.
\end{equation}
The equation for $T_{\mathrm{tr}}$ can be obtained by substituting
$\alpha$ with $2e_1$, $2e_2$, $2e_3$, and then summing up all the three
equations. The result is
\begin{equation} \label{eq:tr_temp}
\frac{\partial T_{\mathrm{tr}}}{\partial t}
  + \sum_{j=1}^3 u_j \frac{\partial T_{\mathrm{tr}}}{\partial x_j}
  + \frac{2}{3\rho R} \sum_{j=1}^3 \left(
    \frac{\partial Q_j}{\partial x_j}
    + \sum_{d=1}^3 \Theta_{jd} \frac{\partial u_d}{\partial x_j}
  \right) = \frac{\mathrm{Pr}}{Z} \cdot \frac{p}{\mu} 
    (T_{\mathrm{eq}} - T_{\mathrm{tr}}),
\end{equation}
where
\begin{equation} \label{eq:Q}
Q_j = 2f_{3e_j,0} + \sum_{d=1}^3 f_{e_j + 2e_d,0}, \quad j = 1,2,3.
\end{equation}
Similarly, if we set $\alpha = 0$ and $k = 1$, then we have
\begin{equation} \label{eq:int_temp}
\frac{\partial T_{\mathrm{int}}}{\partial t}
  + \sum_{j=1}^3 u_j \frac{\partial T_{\mathrm{int}}}{\partial x_j}
  - \frac{2}{\delta \rho R} \sum_{j=1}^3
    \frac{\partial f_{e_j,1}}{\partial x_j}
= \frac{\mathrm{Pr}}{Z} \cdot \frac{p}{\mu}
  (T_{\mathrm{eq}} - T_{\mathrm{int}}).
\end{equation}

\subsection{Truncation and closure with regularization}
Since the moment system \eqref{eq:mnt_eqs} contains an infinite number
of equations and cannot be used for computation, we need to choose a
finite set from them as the governing equations of our method.
However, due to the existence of the last term in the second line of
\eqref{eq:mnt_eqs}, the resulting moment system will be unclosed,
which leads to the ``closure problem'' for the moment method.

We first consider the truncation of the spectral expansion. In
general, we can choose two non-negative integers $M_0 \geqslant 2$ and
$M_1 \geqslant 0$, and use the moment set $\{f_{\alpha,0}\}_{|\alpha|
  \leqslant M_0} \cup \{f_{\alpha,1}\}_{|\alpha| \leqslant M_1}$ as
the finite subset. Such choice well retains the Galilean invariance
since $\{f_{\alpha,0}\}$ and $\{f_{\alpha,1}\}$ only couple with each
other in the collision term. We will postpone the discussion of the
relation between $M_0$ and $M_1$. Below we use $\mathcal{I}$ to denote
the index set such that the set $\{f_{\alpha,k}\}_{(\alpha,k) \in
  \mathcal{I}}$ contains all the moments appearing in the final moment
equations.

Now we are going to make the system closed. The simplest way is to set
$f_{\alpha+e_j,k} = 0$ in \eqref{eq:mnt_eqs} if $(\alpha+e_j,k) \notin
\mathcal{I}$, which leads to the Grad-type moment equations for
polyatomic gases. Mallinger's work \cite{Mallinger} has generalize the
Grad $13$-moment equations to the polyatomic case. Here, we follow
\cite{Reitebuch, Struchtrup2004, NRxx_new} and use the Maxwellian
iteration together with the order of magnitude method to give a more
reasonable closure. The procedure of Maxwellian iteration is
constructed by rearranging \eqref{eq:mnt_eqs} and adding superscripts
representing the iteration steps:
\begin{equation} \label{eq:iter_f}
\begin{split}
f_{\alpha,k}^{(n+1)} & = A_{\alpha,k}^{(n)} - B(\varepsilon) \Bigg\{
  \frac{\partial f_{\alpha,k}^{(n)}}{\partial t}
  + \sum_{d=1}^3
    \frac{\partial u_d}{\partial t} f_{\alpha-e_d,k}^{(n)} \\
& \qquad + \frac{1}{2}
  \frac{\partial (R T_{\mathrm{tr}}^{(n+1)})}{\partial t}
    \sum_{d=1}^3 f_{\alpha-2e_d,k}^{(n)}
  - (m+k) \frac{\partial (R T_{\mathrm{int}}^{(n+1)})}{\partial t}
    f_{\alpha,k-1}^{(n)} \\
& \qquad + \sum_{j=1}^3 \Bigg[ \Bigg(
  R T_{\mathrm{tr}}^{(n+1)}
    \frac{\partial f_{\alpha-e_j,k}^{(n)}}{\partial x_j}
  + u_j \frac{\partial f_{\alpha,k}^{(n)}}{\partial x_j}
  + (\alpha_j + 1)
    \frac{\partial f_{\alpha+e_j,k}^{(n)}}{\partial x_j}
\Bigg) \\
& \qquad \quad + \sum_{d=1}^3 \frac{\partial u_d}{\partial x_j}
  \left(
    R T_{\mathrm{tr}}^{(n+1)} f_{\alpha-e_d-e_j,k}^{(n)}
    + u_j f_{\alpha-e_d,k}^{(n)}
    + (\alpha_j + 1) f_{\alpha-e_d+e_j,k}^{(n)}
  \right) \\
& \qquad \quad + \frac{1}{2}
  \frac{\partial (R T_{\mathrm{tr}}^{(n+1)})}{\partial x_j}
  \sum_{d=1}^3 \left(
    R T_{\mathrm{tr}}^{(n+1)} f_{\alpha-2e_d-e_j,k}^{(n)}
    + u_j f_{\alpha-2e_d,k}^{(n)}
    + (\alpha_j + 1) f_{\alpha-2e_d+e_j,k}^{(n)}
  \right) \\
& \qquad \quad -(m+k)
  \frac{\partial (R T_{\mathrm{int}}^{(n+1)})}{\partial x_j} \left(
    R T_{\mathrm{tr}}^{(n+1)} f_{\alpha-e_j,k-1}^{(n)}
    + u_j f_{\alpha,k-1}^{(n)}
    + (\alpha_j + 1) f_{\alpha+e_j,k-1}^{(n)}
  \right) \Bigg] \Bigg\}, \\
\end{split}
\end{equation}
where $\varepsilon = \mu / (\mathrm{Pr} \cdot p)$ is considered as a
small parameter and $(\alpha,k)$ satisfies
\begin{equation}
(\alpha,k) \in \mathcal{S} := \{\bbN^3 \times \{0,1\} :
  |\alpha| \geqslant 2 \text{ if } k = 0 \text{, }
  |\alpha| \geqslant 1 \text{ if } k = 1 \}.
\end{equation}
In the first line of \eqref{eq:iter_f}, $A_{\alpha,k}^{(n)}$ and
$B(\varepsilon)$ are defined as
\begin{align}
A_{\alpha,k}^{(n)} &= \left\{ \begin{array}{ll}
  \frac{1}{2} \mathrm{Pr} \cdot Z^{-1} \rho
    (R T_{\mathrm{eq}} - R T_{\mathrm{tr}}^{(n+1)}) \delta_{ij},
  & \text{if } \alpha = e_i + e_j, \: k = 0, \\
  G_{\alpha,k}^{(n)}, & \text{other cases}.
\end{array} \right. \\
B(\varepsilon) &= \left\{ \begin{array}{ll}
  \mathrm{Pr} \cdot \varepsilon
    & \text{if } |\alpha| = 2, \: k = 0, \\
\varepsilon, & \text{other cases}.
\end{array} \right.
\end{align}
The reason why the iteration scheme for $|\alpha| = 2$ is special is
that $G_{\alpha,k}$ are functions of $f_{\alpha}$, $|\alpha| = 2$.
Note that $f_{0,0}$($\rho$), $\bu$, $f_{e_j,0}$($\equiv 0$) and
$f_{0,1}$($\equiv 0$) remain invariant during the iteration, and
according to \eqref{eq:tr_temp} and \eqref{eq:int_temp},
$T_{\mathrm{tr}}$ and $T_{\mathrm{int}}$ evolves as
follows:
\begin{align}
\label{eq:iter_T_tr}
T_{\mathrm{tr}}^{(n+1)} &= T_{\mathrm{eq}} - \varepsilon Z \left[
  \frac{\partial T_{\mathrm{tr}}^{(n)}}{\partial t} +
  \sum_{j=1}^3 u_j
    \frac{\partial T_{\mathrm{tr}}^{(n)}}{\partial x_j} +
  \frac{2}{3 \rho R} \sum_{j=1}^3 \left(
    \frac{\partial Q_j^{(n)}}{\partial x_j} +
    \sum_{d=1}^3 \Theta_{jd}^{(n)} \frac{\partial u_d}{\partial x_j}
  \right)
\right], \\
\label{eq:iter_T_int}
T_{\mathrm{int}}^{(n+1)} &= T_{\mathrm{eq}} - \varepsilon Z \left[
  \frac{\partial T_{\mathrm{int}}^{(n)}}{\partial t}
  + \sum_{j=1}^3 u_j
    \frac{\partial T_{\mathrm{int}}^{(n)}}{\partial x_j}
  - \frac{2}{\delta \rho R} \sum_{j=1}^3
    \frac{\partial f_{e_j,1}^{(n)}}{\partial x_j}
\right],
\end{align}
where $T_{\mathrm{eq}}$ also remains invariant during the iteration.
The iteration starts with
\begin{equation}
T_{\mathrm{tr}}^{(0)} = T_{\mathrm{int}}^{(0)} = T_{\mathrm{eq}},
  \qquad f_{0,0}^{(0)} = \rho,
  \qquad f_{\alpha,k}^{(0)} = 0 \text{ for } (\alpha,k) \neq (0,0).
\end{equation}

In order to simplify the notation, we define the following vectors:
\begin{equation}
\bF_{m,k}^{(n)} = (f_{\alpha,k}^{(n)})_{|\alpha| = m}, \quad
\bF_{[m_1,m_2],k}^{(n)} = (\bF_{m_1,k}^{(n)}, \cdots, \bF_{m_2,k}^{(n)}),
  \quad m_1 < m_2.
\end{equation}
and rewrite \eqref{eq:iter_f} and \eqref{eq:iter_T_tr} as
\begin{align}
\label{eq:siter_f}
f_{\alpha,k}^{(n+1)} & = A_{\alpha,k}^{(n)} -
  \varepsilon \boldsymbol{\mathcal{L}}_{\alpha,k}^{(n+1)} \cdot
    \bF_{[|\alpha|-3,|\alpha|+1],k}^{(n)} +
  \varepsilon \boldsymbol{L}_{\alpha,k}^{(n+1)} \cdot
    \bF_{[|\alpha|-1,|\alpha|+1],k-1}^{(n)}, \\
\label{eq:siter_T_tr}
T_{\mathrm{tr}}^{(n+1)} &= T_{\mathrm{eq}} - \varepsilon
  \mathcal{L}(T_{\mathrm{tr}}^{(n)}, \bF_{[2,3],0}^{(n)}), \\
\label{eq:siter_T_int}
T_{\mathrm{int}}^{(n+1)} &= T_{\mathrm{eq}} - \varepsilon
  \mathcal{L}(T_{\mathrm{int}}^{(n)}, \bF_{1,1}^{(n)}),
\end{align}
where $\mathcal{L}(\cdot,\cdot)$ is a linear operator, and
$\boldsymbol{\mathcal{L}}_{\alpha,k}^{(n)}$ is a vector of linear
operators. Each of $\boldsymbol{\mathcal{L}}^{(n)}$'s components has
the following form:
\begin{equation} \label{eq:cal_L}
\sum_{j=1}^3 \sum_{s_1+s_2+s_3+s_4 \leqslant 1} \sum_{r=0}^2
  C_{j,r,\boldsymbol{s}}
  \frac{\partial^{s_1 + s_2} (R T_{\mathrm{tr}}^{(n)})^r}
    {\partial t^{s_1} \partial x_j^{s_2}}
  \frac{\partial^{s_3 + s_4}}{\partial t^{s_3} \partial x_j^{s_4}},
\qquad \boldsymbol{s} \in \{0,1\}^4,
\end{equation}
where $C_{j,r,\boldsymbol{s}} = C_{j,r,\boldsymbol{s}} (\rho, \bu,
\partial / \partial t, \nabla_{\bx})$, which can be considered as a
``constant'' during the iteration. $\boldsymbol{L}_{\alpha,k}^{(n)}$
is a vector, whose components can be expressed as
\begin{equation} \label{eq:L}
\sum_{j=1}^3 \sum_{s_1 + s_2 = 1, s_3 + s_4 \leqslant 1}
  C_{j,\boldsymbol{s}}
  \frac{\partial^{s_1 + s_2} (R T_{\mathrm{int}}^{(n)})}
    {\partial x_j^{s_1} \partial t^{s_2}}
  (R T_{\mathrm{tr}}^{(n)})^{s_3} u_j^{s_4},
\qquad \boldsymbol{s} \in \{0,1\}^4,
\end{equation}
where $C_{j,\boldsymbol{s}}$ are also constants. Now we are ready to
carry out the Maxwellian iteration.

\paragraph{The first step of iteration} In the first step, the
formulae for the translational and internal temperatures can be
written as
\begin{equation} \label{eq:temp}
T_{\mathrm{tr}}^{(1)} = T_{\mathrm{eq}} + \varepsilon S^{(0)},
\quad T_{\mathrm{int}}^{(1)} = T_{\mathrm{eq}} + \varepsilon U^{(0)},
\end{equation}
where $S^{(0)} \sim O(1)$ and $U^{(0)} \sim O(1)$. It is easy to find
\begin{equation} \label{eq:G_0}
G_{\alpha,k}^{(0)} = f_{\alpha,k}^{(0)}, \qquad
  \forall \alpha \in \bbN^3, \quad k = 0,1.
\end{equation}
Thus
\begin{equation}
A_{\alpha,k}^{(0)} = \left\{ \begin{array}{ll}
  -\frac{1}{2} \varepsilon \mathrm{Pr} \cdot Z^{-1} \rho R S^{(0)}, &
    \text{if $\alpha = 2e_i$, $k = 0$}, \\
  0, & \text{other cases}.
\end{array} \right.
\end{equation}
Now, it can be easily deduced from \eqref{eq:siter_f} that
$f_{\alpha,k}^{(1)}$ is nonzero if and only if $0 \in [|\alpha|-3,
|\alpha|+1]$, $k = 0$ or $0 \in [|\alpha|-1, |\alpha|+1]$, $k = 1$.
Precisely, we have
\begin{equation} \label{eq:f_1}
f_{\alpha,k}^{(1)} \sim \left\{ \begin{array}{ll}
  O(\varepsilon), &
    (\alpha,k) \in \mathcal{S} \text{ and }
    |\alpha| \leqslant 3 \text{, } k = 0, \\
  O(\varepsilon), &
    (\alpha,k) \in \mathcal{S} \text{ and }
    |\alpha| = 1 \text{, } k = 1, \\
  0, & \text{other cases for } (\alpha,k) \in \mathcal{S},
\end{array} \right.
\end{equation}
and
\begin{equation}
\bF_{[2,3],0}^{(1)} = \bF_{[2,3],0}^{(0)} + \varepsilon \bH_{[2,3],0}^{(0)},
\quad \bF_{1,1}^{(1)} = \bF_{1,1}^{(0)} + \varepsilon \bH_{1,1}^{(0)},
\end{equation}
where $\bH_{[2,3],k}^{(0)}$ and $\bH_{1,1}^{(0)}$ has an order of
magnitude $O(1)$. The meaning of the subscripts of $\bH$ is the same
as $\bF$.

Now let us consider $G_{\alpha,k}^{(1)}$. $G_{0,0}$, $G_{0,1}$ and
$G_{\alpha,k}$ with odd $|\alpha|$ keep invariant during the
iteration. For $G_{\alpha,0}$ with $|\alpha| \geqslant 2$,
\eqref{eq:G_recur} gives
\begin{equation} \label{eq:G_L}
G_{\alpha,0}^{(n)} = \sum_{j=1}^3 L_{\alpha,j} \left(
  T_{\mathrm{tr}}^{(n)} - T_{\mathrm{eq}},
  \bF_{2,0}^{(n)}
\right) G_{\alpha-e_i-e_j, 0}^{(n)},
\end{equation}
where $L_{\alpha,j}(\cdot, \cdot)$ is a linear function independent of
$\varepsilon$. Using \eqref{eq:f_1}, \eqref{eq:temp} and $G_{0,0} =
\rho \sim O(1)$, we can easily get
\begin{equation}
G_{\alpha,0}^{(1)} \sim O(\varepsilon^{|\alpha|/2}), \qquad
  |\alpha| \text{ is even}.
\end{equation}
Using \eqref{eq:C}, we have
\begin{equation}
G_{\alpha,1}^{(1)} = \frac{\delta R}{2Z}
  \varepsilon U^{(0)} G_{\alpha,0}^{(1)}
  \sim O(\varepsilon^{|\alpha|/2+1}), \qquad
  |\alpha| \text{ is even}.
\end{equation}

\paragraph{The general progress} In general case, we have
\begin{equation} \label{eq:f_n}
f_{\alpha,k}^{(n+1)} \sim \left\{ \begin{array}{ll}
  O(\varepsilon^{\lceil |\alpha|/3 \rceil}), &
    (\alpha,k) \in \mathcal{S} \text{ and }
    |\alpha| \leqslant 3(n+1) \text{, } k = 0, \\
  o(\varepsilon^{\lceil |\alpha|/3 \rceil}), &
    (\alpha,k) \in \mathcal{S} \text{ and }
    |\alpha| > 3(n+1) \text{, } k = 0, \\
  O(\varepsilon^{\lceil (|\alpha| + 2) / 3 \rceil}), &
    (\alpha,k) \in \mathcal{S} \text{ and }
    |\alpha| \leqslant 3n+1 \text{, } k = 1, \\
  o(\varepsilon^{\lceil (|\alpha| + 2) / 3 \rceil}), &
    (\alpha,k) \in \mathcal{S} \text{ and }
    |\alpha| > 3n+1 \text{, } k = 1,
\end{array} \right.
\end{equation}
and
\begin{align}
\label{eq:F_0}
\bF_{[2,3(n+1)],0}^{(n+1)} &= \bF_{[2,3(n+1)],0}^{(n)} +
  \varepsilon^{n+1} \bH_{[2,3(n+1)],0}^{(n)},\\
\label{eq:F_1}
\bF_{[1,3n+1],1}^{(n+1)} &= \bF_{[1,3n+1],1}^{(n)} +
  \varepsilon^{n+1} \bH_{[1,3n+1],1}^{(n)},\\
\label{eq:T_tr_n}
T_{\mathrm{tr}}^{(n+1)} &= T_{\mathrm{tr}}^{(n)} + 
  \varepsilon^{n+1} S^{(n)}, \\
\label{eq:T_int_n}
T_{\mathrm{int}}^{(n+1)} &= T_{\mathrm{int}}^{(n)} + 
  \varepsilon^{n+1} U^{(n)},
\end{align}
where $\bH_{[m_1,m_2],k}^{(n)} \sim O(1)$, $S^{(n)} \sim O(1)$ and
$U^{(n)} \sim O(1)$. \eqref{eq:f_n}---\eqref{eq:T_int_n} can be
validated by induction.  Through the derivation in the last paragraph,
we have known that \eqref{eq:f_n}---\eqref{eq:T_int_n} hold for $n =
0$. Now we suppose \eqref{eq:f_n}---\eqref{eq:T_int_n} hold for $n -
1$. Then \eqref {eq:G_L} gives
\begin{equation}
G_{\alpha,0}^{(n)} = \sum_{j=1}^3 \left[
  L_{\alpha,j} \left(
    T_{\mathrm{tr}}^{(n-1)} - T_{\mathrm{eq}}, \bF_{2,0}^{(n-1)}
  \right) + \varepsilon^{n} L_{\alpha,j} \left(
    S^{(n-1)}, \bH_{2,0}^{(n-1)}
  \right)
\right] G_{\alpha-e_i-e_j,0}^{(n)}.
\end{equation}
Now using $G_{0,0}^{(n)} = G_{0,0}^{(n-1)} = \rho$ and \eqref{eq:G_L}
with $n$ replaced by $n-1$, simple induction gives
\begin{equation} \label{eq:G_alpha_0_diff}
G_{\alpha,0}^{(n)} - G_{\alpha,0}^{(n-1)} \sim
  O(\varepsilon^{n+|\alpha|/2-1}), \qquad
|\alpha| \geqslant 2 \text{ and } |\alpha| \text{ is even}.
\end{equation}
For $G_{\alpha,1}^{(n)}$, we have
\begin{equation} \label{eq:G_alpha_1_diff}
\begin{split}
G_{\alpha,1}^{(n)} - G_{\alpha,1}^{(n-1)} &=
  C^{(n)} G_{\alpha,0}^{(n)} - C^{(n-1)} G_{\alpha,0}^{(n-1)} \\
&= \left[
  C^{(n-1)} + \frac{\delta R}{2Z} \varepsilon^n U^{(n-1)}
\right] G_{\alpha,0}^{(n)} - C^{(n-1)} G_{\alpha,0}^{(n-1)} \\
&= C^{(n-1)} (G_{\alpha,0}^{(n)} - G_{\alpha,0}^{(n-1)})
  + \frac{\delta R}{2Z} \varepsilon^n U^{(n-1)} G_{\alpha,0}^{(n)} \\
&\sim O(\varepsilon) \cdot O(\varepsilon^{n+|\alpha|/2-1})
  + \varepsilon^n \cdot O(1) \cdot O(\varepsilon^{|\alpha|/2}) \\
&\sim O(\varepsilon^{n+|\alpha|/2}), \qquad
  |\alpha| \geqslant 2 \text{ and } |\alpha| \text{ is even}.
\end{split}
\end{equation}
Similarly, by \eqref{eq:siter_T_tr}, we have
\begin{equation} \label{eq:T_tr_diff}
\begin{split}
T_{\mathrm{tr}}^{(n+1)} &= T_{\mathrm{eq}} - \varepsilon \left[
  \mathcal{L} \left(
    T_{\mathrm{tr}}^{(n-1)}, \bF_{[2,3],0}^{(n-1)}
  \right) + \varepsilon^{n} \mathcal{L} \left(
    S^{(n-1)}, \bH_{[2,3],0}^{(n-1)}
  \right)
\right] \\
&= T_{\mathrm{tr}}^{(n)} -
  \varepsilon^{n+1} \mathcal{L} \left(
    S^{(n-1)}, \bH_{[2,3],0}^{(n-1)}
  \right).
\end{split}
\end{equation}
Defining $S^{(n)} = -\mathcal{L} \left( S^{(n-1)},
\bH_{[2,3],0}^{(n-1)} \right)$ gives \eqref{eq:T_tr_n}. The validation
of \eqref{eq:T_int_n} is almost the same. \eqref{eq:G_alpha_0_diff},
\eqref{eq:G_alpha_1_diff} and \eqref {eq:T_tr_diff} imply that
$A_{\alpha,k}^{(n)} - A_{\alpha,k}^{(n-1)}$ is never greater than
$O(\varepsilon^{n+1})$.

Now it remains to consider $f_{\alpha,k}^{(n+1)}$. Similar as \eqref
{eq:G_alpha_1_diff}, the equations \eqref{eq:cal_L}, \eqref{eq:L}
and \eqref{eq:T_tr_diff} show
\begin{gather}
\label{eq:cal_L_alpha_k}
\boldsymbol{\mathcal{L}}_{\alpha,k}^{(n+1)} - 
  \boldsymbol{\mathcal{L}}_{\alpha,k}^{(n)} \sim O(\varepsilon^{n+1}),
\qquad \boldsymbol{\mathcal{L}}_{\alpha,k}^{(n+1)} \sim O(1), \\
\label{eq:L_alpha_k}
\boldsymbol{L}_{\alpha,k}^{(n+1)} - \boldsymbol{L}_{\alpha,k}^{(n)}
  \sim O(\varepsilon^{n+1}),
\qquad \boldsymbol{L}_{\alpha,k}^{(n+1)} \sim O(1),
\end{gather}
Using the assumptions of the induction, one has
\begin{equation}
\begin{split}
f_{\alpha,0}^{(n+1)} & = A_{\alpha,0}^{(n)} -
  \varepsilon \boldsymbol{\mathcal{L}}_{\alpha,0}^{(n+1)} \cdot
    \bF_{[|\alpha|-3,|\alpha|+1],0}^{(n)} \\
& \sim \left\{ \begin{array}{ll}
  O(\varepsilon^{|\alpha|/2}) + \varepsilon O(1) \cdot
  O(\varepsilon^{\lceil (|\alpha| - 3) / 3 \rceil})
    \sim O(\varepsilon^{\lceil |\alpha| / 3 \rceil}),
  & |\alpha| \leqslant 3(n+1), \\
  O(\varepsilon^{|\alpha|/2}) + \varepsilon O(1) \cdot
  o(\varepsilon^{\lceil (|\alpha| - 3) / 3 \rceil})
    \sim o(\varepsilon^{\lceil |\alpha| / 3 \rceil}), 
  & |\alpha| > 3(n+1),
\end{array} \right.
\end{split}
\end{equation}
and
\begin{equation}
\begin{split}
f_{\alpha,1}^{(n+1)} & = A_{\alpha,1}^{(n)} -
  \varepsilon \boldsymbol{\mathcal{L}}_{\alpha,1}^{(n+1)} \cdot
    \bF_{[|\alpha|-3,|\alpha|+1],1}^{(n)} +
  \varepsilon \boldsymbol{L}_{\alpha,1}^{(n+1)} \cdot
    \bF_{[|\alpha|-1,|\alpha|+1],0}^{(n)} \\
& \sim \left\{ \begin{array}{ll}
  O(\varepsilon^{|\alpha|/2+1}) +
  \varepsilon O(\varepsilon^{\lceil (|\alpha| - 1) / 3 \rceil}) +
  \varepsilon O(\varepsilon^{\lceil (|\alpha| - 1) / 3 \rceil})
    \sim O(\varepsilon^{\lceil (|\alpha|+2) / 3 \rceil}),
  & |\alpha| \leqslant 3n+1, \\
  O(\varepsilon^{|\alpha|/2+1}) +
  \varepsilon o(\varepsilon^{\lceil (|\alpha| - 1) / 3 \rceil})+
  \varepsilon o(\varepsilon^{\lceil (|\alpha| - 1) / 3 \rceil})
    \sim o(\varepsilon^{\lceil (|\alpha|+2) / 3 \rceil}), 
  & |\alpha| > 3n+1.
\end{array} \right.
\end{split}
\end{equation}
This gives \eqref{eq:f_n}. The validation of \eqref{eq:F_0} and
\eqref{eq:F_1} needs
\begin{equation}
\begin{split}
f_{\alpha,k}^{(n+1)} - f_{\alpha,k}^{(n)} &=
  A_{\alpha,k}^{(n)} - A_{\alpha,k}^{(n-1)} -
  \varepsilon (\boldsymbol{\mathcal{L}}_{\alpha,k}^{(n+1)} \cdot
    \bF_{[|\alpha|-3,|\alpha|+1],k}^{(n)} -
  \boldsymbol{\mathcal{L}}_{\alpha,k}^{(n+1)} \cdot
    \bF_{[|\alpha|-3,|\alpha|+1],k}^{(n)}) \\
&\phantom{{} = G_{\alpha,k}^{(n)} - G_{\alpha,k}^{(n-1)}} +
  \varepsilon (\boldsymbol{L}_{\alpha,k}^{(n+1)} \cdot
    \bF_{[|\alpha|-1,|\alpha|+1],k-1}^{(n)} -
  \boldsymbol{L}_{\alpha,k}^{(n+1)} \cdot
    \bF_{[|\alpha|-1,|\alpha|+1],k-1}^{(n)}) \\
&= A_{\alpha,k}^{(n)} - A_{\alpha,k}^{(n-1)} -
  \varepsilon \boldsymbol{\mathcal{L}}_{\alpha,k}^{(n+1)} \cdot
    (\bF_{[|\alpha|-3,|\alpha|+1],k}^{(n)} -
    \bF_{[|\alpha|-3,|\alpha|+1],k}^{(n-1)}) \\
&\phantom{{} = G_{\alpha,k}^{(n)} - G_{\alpha,k}^{(n-1)}} -
  \varepsilon (\boldsymbol{\mathcal{L}}_{\alpha,k}^{(n+1)} -
    \boldsymbol{\mathcal{L}}_{\alpha,k}^{(n)}) \cdot
    \bF_{[|\alpha|-3,|\alpha|+1],k}^{(n-1)}, \\
&\phantom{{} = G_{\alpha,k}^{(n)} - G_{\alpha,k}^{(n-1)}} +
  \varepsilon \boldsymbol{L}_{\alpha,k}^{(n)} \cdot
    (\bF_{[|\alpha|-1,|\alpha|+1],k-1}^{(n)} -
    \bF_{[|\alpha|-1,|\alpha|+1],k-1}^{(n-1)}) \\
&\phantom{{} = G_{\alpha,k}^{(n)} - G_{\alpha,k}^{(n-1)}} +
  \varepsilon (\boldsymbol{L}_{\alpha,k}^{(n+1)} -
    \boldsymbol{L}_{\alpha,k}^{(n)}) \cdot
    \bF_{[|\alpha|-1,|\alpha|+1],k-1}^{(n-1)},
\end{split}
\end{equation}
and then \eqref{eq:G_alpha_0_diff}, \eqref{eq:G_alpha_1_diff},
\eqref{eq:cal_L_alpha_k} and \eqref{eq:L_alpha_k} show that
\begin{equation}
f_{\alpha,k}^{(n+1)} - f_{\alpha,k}^{(n)} \sim O(\varepsilon^{n+1}).
\end{equation}

The equations \eqref{eq:f_n}---\eqref{eq:F_1} tell us that
$f_{\alpha,k}^{(n)}$ is never greater than $O(\varepsilon^{\lceil
(|\alpha| + 2k) /3 \rceil})$ for arbitrary $n$, and the leading
order of $f_{\alpha,k}$ appears at the $\lceil (|\alpha| + 2k) / 3
\rceil$-th iteration step, and never changes later. Based on these
results, we can remove some high order terms in the moment equations
\eqref{eq:mnt_eqs}, and the remaining part is the formula for the
regularization term. Precisely, we have
\begin{equation} \label{eq:hot_collected}
\begin{split}
& \sum_{d=1}^3 \frac{\partial u_d}{\partial t} f_{\alpha-e_d,k}
  + \frac{1}{2} \frac{\partial (R T_{\mathrm{tr}})}{\partial t}
    \sum_{d=1}^3 f_{\alpha-2e_d,k}
  + \sum_{j=1}^3 R T_{\mathrm{tr}}
    \frac{\partial f_{\alpha-e_j,k}}{\partial x_j} \\
& \qquad - (m + k) \frac{\partial (R T_{\mathrm{int}})}{\partial t}
    f_{\alpha, k-1}
  + \sum_{j=1}^3 \sum_{d=1}^3 \frac{\partial u_d}{\partial x_j} \left(
    R T_{\mathrm{tr}} f_{\alpha-e_d-e_j,k} + u_j f_{\alpha-e_d,k}
  \right) \\
& \qquad + \frac{1}{2} \sum_{j=1}^3
  \frac{\partial(R T_{\mathrm{tr}})}{\partial x_j} \sum_{d=1}^3 \left(
    R T_{\mathrm{tr}} f_{\alpha-2e_d-e_j,k}
    + u_j f_{\alpha-2e_d,k}
    + (\alpha_j + 1) f_{\alpha-2e_d+e_j,k}
  \right) \\
& \qquad - (m + k) \sum_{j=1}^3
  \frac{\partial (R T_{\mathrm{int}})}{\partial x_j}
  (R T_{\mathrm{tr}} f_{\alpha-e_j,k-1} + u_j f_{\alpha,k-1}
    + (\alpha_j + 1) f_{\alpha+e_j,k-1}) \\
& \hspace{275pt} = \frac{1}{\varepsilon} (G_{\alpha,k} - f_{\alpha,k})
  + h.o.t., \end{split}
\end{equation}
where $h.o.t.$ stands for high order terms. The equations
\eqref{eq:u}, \eqref{eq:tr_temp} and \eqref{eq:int_temp} can be used
to make \eqref{eq:hot_collected} more compact. Noting that the terms
containing $f_{e_i+e_j,0}$ or $f_{e_j,1}$ can be regarded as a high
order term in \eqref{eq:Theta}, we can write \eqref{eq:hot_collected}
as
\begin{equation} \label{eq:f_alpha_k}
\begin{split}
f_{\alpha,k} &= \varepsilon \Bigg[
  \sum_{j=1}^3 \left(
    \frac{1}{\rho}
      \frac{\partial (\rho R T_{\mathrm{tr}})}{\partial x_j}
      f_{\alpha-e_j,k} -
    R T_{\mathrm{tr}} \frac{\partial f_{\alpha-e_j,k}}{\partial x_j}
  \right) + \frac{1}{3} R T_{\mathrm{tr}} \left(
    \sum_{j=1}^3 \frac{\partial u_j}{\partial x_j}
  \right) \sum_{d=1}^3 f_{\alpha-2e_d,k} \\
& \quad \quad \quad \quad - \sum_{j=1}^3 \sum_{d=1}^3
  \frac{1}{2} \frac{\partial (R T_{\mathrm{tr}})}{\partial x_j}
    (R T_{\mathrm{tr}} f_{\alpha-2e_d-e_j,k} +
     (\alpha_j + 1) f_{\alpha-2e_d+e_j,k}) \\
& \quad \quad \quad \quad + (m + k) \sum_{j=1}^3
  \frac{\partial (R T_{\mathrm{int}})}{\partial x_j}
    (R T_{\mathrm{tr}} f_{\alpha-e_j,k-1} +
     (\alpha_j + 1) f_{\alpha+e_j,k-1}) \\
& \quad \quad \quad \quad \quad \quad - \sum_{j=1}^3 \sum_{d=1}^3
  \frac{\partial u_d}{\partial x_j}
    R T_{\mathrm{tr}} f_{\alpha-e_d-e_j,k} \Bigg]
  + G_{\alpha,k} + h.o.t..
\end{split}
\end{equation}
If $|\alpha| = 2$ and $k = 0$, the above equation becomes
\begin{equation}
\begin{split}
f_{e_i+e_j,0} &= \varepsilon \left[
  \frac{1}{3} \delta_{ij} \rho R T_{\mathrm{tr}}
    \sum_{d=1}^3 \frac{\partial u_d}{\partial x_d} -
  \frac{1}{1 + \delta_{ij}} \rho R T_{\mathrm{tr}} \left(
    \frac{\partial u_i}{\partial x_j} +
    \frac{\partial u_j}{\partial x_i}
  \right)
\right] \\
& \qquad + (1 - \mathrm{Pr}^{-1}) f_{e_i+e_j,0} +
  \frac{1}{2} \delta_{ij} Z^{-1}
    \rho (R T_{\mathrm{eq}} - R T_{\mathrm{tr}}) + h.o.t..
\end{split}
\end{equation}
With \eqref{eq:Theta}, it is reformulated as
\begin{equation} \label{eq:NS}
(1 + \delta_{ij}) f_{e_i+e_j,0} =
  -2\mathrm{Pr} \cdot \varepsilon \rho R T_{\mathrm{tr}}
    \frac{\partial u_{\langle i}}{\partial x_{j\rangle}}
  + \delta_{ij} \mathrm{Pr} \cdot Z^{-1}
    \rho (R T_{\mathrm{eq}} - R T_{\mathrm{tr}}) + h.o.t.,
\end{equation}
which is similar as the Navier-Stokes law. Analogously, we can get the
following relations similar as the Fourier's law:
\begin{equation} \label{eq:Fourier}
\begin{split}
Q_j & = -\frac{5}{2} \varepsilon \rho R T_{\mathrm{tr}}
  \frac{\partial (R T_{\mathrm{tr}})}{\partial x_j} + h.o.t., \\
f_{e_j,1} &= \frac{\delta}{2} \varepsilon \rho R T_{\mathrm{tr}}
  \frac{\partial (R T_{\mathrm{int}})}{\partial x_j} + h.o.t..
\end{split}
\end{equation}
\eqref{eq:NS} and \eqref{eq:Fourier} can be used to eliminate most
partial derivatives in \eqref{eq:f_alpha_k}. Direct substitution gives
the following result for $|\alpha| \geqslant 3$:
\begin{equation} \label{eq:regularization}
\begin{split}
f_{\alpha,k} &= \varepsilon \sum_{j=1}^3 \left(
  \frac{1}{\rho} \frac{\partial(\rho R T_{\mathrm{tr}})}{\partial x_j}
    f_{\alpha-e_j,k}
  - R T_{\mathrm{tr}}
    \frac{\partial f_{\alpha-e_j,k}}{\partial x_j}
\right) \\
& \qquad +\frac{1}{2 \mathrm{Pr} \cdot \rho} \sum_{j=1}^3 \sum_{d=1}^3
  (1 + \delta_{jd}) f_{e_j+e_d,0} f_{\alpha-e_d-e_j,k} \\
& \qquad -\frac{1}{2} Z^{-1} (R T_{\mathrm{eq}} - R T_{\mathrm{tr}})
  \sum_{j=1}^3 f_{\alpha-e_d-e_j,k} \\
& \qquad + \frac{1}{5\rho R T_{\mathrm{tr}}} \sum_{j=1}^3 \sum_{d=1}^3
  Q_j [R T_{\mathrm{tr}} f_{\alpha-2e_d-e_j,k} +
    (\alpha_j + 1) f_{\alpha-2e_d+e_j,k}] \\
& \qquad + \frac{1}{\rho R T_{\mathrm{tr}}} \sum_{j=1}^3 f_{e_j,1}
  [R T_{\mathrm{tr}} f_{\alpha-e_j,k-1} +
   (\alpha_j + 1) f_{\alpha+e_j,k-1}] + h.o.t..
\end{split}
\end{equation}
Neglecting the high order terms and applying the result for $|\alpha|
= M_0 + 1$, $k = 0$ and $|\alpha| = M_1 + 1$, $k = 1$, we carry out a
reasonable approximation to the moments of the corresponding orders.
This completes the closure of the moment system.

There is still one question about the relation between $M_0$ and $M_1$
remaining. Since when $|\alpha| \geqslant 1$, $f_{\alpha,1}$ always
has the same order of magnitude as $f_{\alpha+e_i+e_j,0}$, it is
natural to choose $M_0 = M_1 + 2$. Thus the total number of moments is
$(M_0 + 1)(M_0^2 + 2 M_0 + 3) / 3$.

\subsection{Linearization of the regularization term}
The expression \eqref{eq:regularization} is rather complicated and is
inconvenient for numerical implementation. In \cite{Cai}, we have used
the technique of linearization to simplify the regularization
term. Here, the same idea will be applied to derive a simplified
regularization term.

We consider a local problem where the distribution function is around
the Maxwellian and the variations of the density, velocity and
temperature are small. Thus we have the local expansions with a small
parameter $\epsilon$ indicating the magnitude of perturbation around
the Maxwellian as
\begin{equation} \label{eq:perturb}
\begin{gathered}
\rho = \rho_0 (1 + \epsilon \hat{\rho}), \quad
\bu = \bu_0 + \epsilon \sqrt{R T_0} \hat{\bu}, \quad
T_{\mathrm{eq}} = T_0 (1 + \epsilon \hat{T}_{\mathrm{eq}}), \quad
T_{\mathrm{tr}} = T_0 (1 + \epsilon \hat{T}_{\mathrm{tr}}), \\
\bx = L \epsilon \hat{\bx}, \quad
\mu = L \rho_0 \sqrt{R T_0} \epsilon \hat{\mu}, \quad
f_{\alpha,k} = \rho_0 (R T_0)^{|\alpha|/2 + k} \epsilon
  \hat{f}_{\alpha,k} \text{\ \ for\ \ } (\alpha,k) \neq (0,0),
\end{gathered}
\end{equation}
where $\rho_0$, $\bu_0$ and $T_0$ are the reference density, velocity
and temperature, respectively, the variables with hats $\hat{\cdot}$
are dimensionless variables with magnitudes $O(1)$ and $L$ is the
characteristic length. The equations \eqref{eq:pressure} and
\eqref{eq:Q} show that
\begin{equation} \label{eq:perturb_p_Q}
p = \rho_0 R T_0 (1 + \epsilon \hat{p}), \qquad
Q_j = \rho_0 (R T_0)^{3/2} \epsilon \hat{Q}_j, \quad j = 1,2,3,
\end{equation}
where $\hat{p}$ and $\hat{Q}_j$ are also $O(1)$ dimensionless
variables. Now we put \eqref{eq:perturb} and \eqref{eq:perturb_p_Q}
into \eqref{eq:regularization}. Eliminating the constants on both
sides, we get
\begin{equation}
\begin{split}
\hat{f}_{\alpha,k}
  &= \frac{\hat{\mu}}{\mathrm{Pr} (1 + \epsilon \hat{p})}
  \sum_{j=1}^3 \left(
    \frac{\epsilon R}{1 + \epsilon \hat{\rho}}
      \frac{\partial(\hat{\rho} + \hat{T}_{\mathrm{tr}} +
        \epsilon \hat{\rho} \hat{T}_{\mathrm{tr}})}
        {\partial \hat{x}_j}
      \hat{f}_{\alpha-e_j,k}
    - R (1 + \epsilon \hat{T}_{\mathrm{tr}})
      \frac{\partial \hat{f}_{\alpha-e_j,k}}{\partial \hat{x}_j}
  \right) \\
& \qquad +\frac{\epsilon}{2 \mathrm{Pr} (1 + \epsilon \hat{\rho})}
  \sum_{j=1}^3 \sum_{d=1}^3
    (1 + \delta_{jd}) \hat{f}_{e_j+e_d,0} \hat{f}_{\alpha-e_d-e_j,k}\\
& \qquad -\frac{1}{2} \epsilon Z^{-1}
  (R \hat{T}_{\mathrm{eq}} - R \hat{T}_{\mathrm{tr}})
    \sum_{j=1}^3 \hat{f}_{\alpha-e_d-e_j,k} \\
& \qquad +
  \frac{\epsilon}
    {5(1 + \epsilon \hat{\rho}) R (1 + \epsilon \hat{T}_{\mathrm{tr}})}
  \sum_{j=1}^3 \sum_{d=1}^3 \hat{Q}_j
    [R (1 + \epsilon \hat{T}_{\mathrm{tr}}) \hat{f}_{\alpha-2e_d-e_j,k} +
      (\alpha_j + 1) \hat{f}_{\alpha-2e_d+e_j,k}] \\
& \qquad + \frac{\epsilon}
    {(1 + \epsilon \hat{\rho}) R (1 + \epsilon \hat{T}_{\mathrm{tr}})}
  \sum_{j=1}^3 \hat{f}_{e_j,1}
    [R (1 + \epsilon \hat{T}_{\mathrm{tr}}) \hat{f}_{\alpha-e_j,k-1} +
      (\alpha_j + 1) \hat{f}_{\alpha+e_j,k-1}].
\end{split}
\end{equation}
Reserving only leading order terms on the right hand side, one has
\begin{equation}
\hat{f}_{\alpha,k} \approx
  -\frac{\hat{\mu}}{\mathrm{Pr} (1 + \epsilon \hat{p})} \cdot
    R(1 + \epsilon \hat{T}_{\mathrm{tr}}) \sum_{j=1}^3
      \frac{\partial \hat{f}_{\alpha - e_j,k}}{\partial \hat{x}_j}.
\end{equation}
Using \eqref{eq:perturb} and \eqref{eq:perturb_p_Q} again, we get a
simple approximation of $f_{\alpha,k}$:
\begin{equation} \label{eq:linear_reg}
f_{\alpha,k} \approx -\frac{\mu}{\mathrm{Pr} \cdot p} \cdot
  R T_{\mathrm{tr}} \sum_{j=1}^3
    \frac{\partial f_{\alpha - e_j,k}}{\partial x_j}.
\end{equation}
This approximation is to be applied to $|\alpha| = M_k + 1$ and used
in our numerical method.
\begin{remark}
  The linearization may cause the loss of accuracy in the moment
  method.  However, since the {\NRxx} method is applicable up to
  arbitrary order of moments, the loss of accuracy can be got back by
  increasing the highest order in the system by $1$. An important
  advantage of the regularization term \eqref{eq:linear_reg} is to
  smooth the profile of macroscopic variables such that the unphysical
  subshocks can be eliminated (see e.g.  \cite{Grad1952,
    Struchtrup}). We will find in the numerical experiments that the
  regularization term \eqref{eq:linear_reg} actually acts as a
  diffusion.
\end{remark}

\subsection{Numerical method}
The framework of the numerical scheme for the {\NRxx} method in the
polyatomic case is generally the same as that in the monatomic case.
The fractional step method is utilized to treat convection term and
collision term separately. For the convection term, the finite volume
method with the HLL numerical flux is employed. Below we consider the
one-dimensional case and suppose a uniform spatial grid with grid size
$\Delta x$ is used. For an arbitrary quantity $\psi$, the symbol
$\psi_i^n$ denotes the average value of $\psi$ on the $i$-th grid at
the $n$-th time step. Then the whole algorithm is outlined as follows:
\begin{enumerate}
\setlength\itemsep{0cm}
\item Let $n$ be zero and set $f_i^n(\bxi,I)$ to be the initial value.
\item \label{step:reconstruct}%
  For each $i$, apply the technique of linear reconstruction to
  determine the distributions on both boundaries of the $i$-th grid.
  The results are denoted as $f_{i-1/2}^{n+}(\bxi,I)$ and
  $f_{i+1/2}^{n-}(\bxi,I)$.
\item For each reconstructed distribution function, use
  \eqref{eq:linear_reg} for $|\alpha| = M_k + 1$ to approximate the
  truncated moments.
\item Use CFL condition to determine the time step $\Delta t$:
  \begin{equation}
  \Delta t \max_i \left\{
    \frac{|\bu|_i^n + C(M_0) \sqrt{(R T_{\mathrm{tr}})_i^n}}{\Delta x}
      +\frac{2(M+1)}{(\Delta x)^2} (\varepsilon R T_{\mathrm{tr}})_i^n
  \right\}
  \leqslant \frac{1}{2} \mathit{CFL},
  \end{equation}
  where $C(M_0)$ is the maximal root of the Hermite polynomial
  $\He_{M_0}(x)$, and $\mathit{CFL}$ is a specified Courant number
  between $0$ and $1$.
\item \label{step:convection}%
  Solve the convection part with the HLL scheme:
  \begin{equation} \label{eq:FVM}
  f_i^{n*}(\bxi, I) = f_i^n(\bxi, I) - \frac{\Delta t}{\Delta x}
    [G_{i+1/2}^n(\bxi, I) - G_{i-1/2}^n(\bxi, I)],
  \end{equation}
  where
  \begin{equation} \label{eq:flux}
  G_{i+1/2}^n(\bxi) = \left\{ \begin{array}{ll}
    \xi_1 f_{i+1/2}^{n-}(\bxi, I), &
      0 \leqslant \lambda_{i+1/2}^{n-}, \\[5pt]
    \dfrac{\lambda_{i+1/2}^{n+} \xi_1 f_{i+1/2}^{n-}(\bxi, I) -
      \lambda_{i+1/2}^{n-} \xi_1 f_{i+1/2}^{n+}(\bxi, I)}
      {\lambda_{i+1/2}^{n+} - \lambda_{i+1/2}^{n-}} \\[15pt]
    \qquad + \dfrac{\lambda_{i+1/2}^{n-} \lambda_{i+1/2}^{n+}
        [f_{i+1/2}^{n+}(\bxi, I) - f_{i+1/2}^{n-}(\bxi, I)]}
      {\lambda_{i+1/2}^{n+} - \lambda_{i+1/2}^{n-}}, &
      \lambda_{i+1/2}^{n-} < 0 < \lambda_{i+1/2}^{n+}, \\[15pt]
    \xi_1 f_{i+1/2}^{n+}(\bxi, I), & 0 \geqslant \lambda_{i+1/2}^{n+},
  \end{array} \right.
  \end{equation}
  and
  \begin{equation}
  \begin{aligned}
  \lambda_{i+1/2}^{n-} &= \min\left\{
    (u_1)_{i+1/2}^{n-} - C(M_0) \sqrt{(R T_{\mathrm{tr}})_{i+1/2}^{n-}},
    \: (u_1)_{i+1/2}^{n+} - C(M_0) \sqrt{(R T_{\mathrm{tr}})_{i+1/2}^{n+}}
  \right\}, \\
  \lambda_{i+1/2}^{n+} &= \max\left\{
    (u_1)_{i+1/2}^{n-} + C(M_0) \sqrt{(R T_{\mathrm{tr}})_{i+1/2}^{n-}},
    \: (u_1)_{i+1/2}^{n+} + C(M_0) \sqrt{(R T_{\mathrm{tr}})_{i+1/2}^{n+}}
  \right\}.
  \end{aligned}
  \end{equation}
\item \label{step:collision}%
  Solve the collision-only equation by $\Delta t$, and the result is
  denoted as $f_i^{n+1}(\bxi, I)$.
\item Increase $n$ by $1$ and return to Step \ref{step:reconstruct}.
\end{enumerate}
The details of linear reconstruction and the numerical approximation
of the truncated moments are almost the same as the monatomic case,
and we refer the readers to \cite{Cai} for our implementation. Here
only Step \ref{step:convection} and Step \ref{step:collision} are
expanded in the following subsections.

\subsubsection{The convection step}
In \cite{NRxx}, we have mentioned that two operations are needed to
accomplish the HLL scheme. One is the calculation of $\xi_j f$ for $f$
represented by \eqref{eq:expansion}, and the other is the linear
operation on distribution functions. The former is straightforward:
\begin{equation} \label{eq:times_xi}
\begin{split}
& \phantom{={}}
  \xi_j f(\bxi, I) = \left( \sqrt{R T_{\mathrm{tr}}} v_j + u_j \right)
  \sum_{\alpha \in \bbN^3} \sum_{k \in \bbN} f_{\alpha,k}
  \psi_{\alpha,k,T_{\mathrm{tr}},T_{\mathrm{int}}} (\bv, J) \\
&= \sum_{\alpha \in \bbN^3} \sum_{k \in \bbN} f_{\alpha,k} \left[
  R T_{\mathrm{tr}}
    \psi_{\alpha+e_j,k,T_{\mathrm{tr}},T_{\mathrm{int}}} (\bv, J)
  + u_j \psi_{\alpha,k,T_{\mathrm{tr}},T_{\mathrm{int}}} (\bv, J)
  + \alpha_j
    \psi_{\alpha-e_j,k,T_{\mathrm{tr}},T_{\mathrm{int}}} (\bv, J)
\right],
\end{split}
\end{equation}
where the recursion relation of Hermite polynomials is used, and
\begin{equation}
\bv = (\bxi - \bu) / \sqrt{R T_{\mathrm{tr}}}, \qquad
  J = I^{2/\delta} / (R T_{\mathrm{int}}).
\end{equation}

In order to make linear operations on distribution functions
applicable, we have to solve the following problem:
\begin{quote} \it%
For $f(\bxi, I)$ represented by \eqref{eq:expansion}, find a series of
coefficients $f_{\alpha,k}'$, such that 
\begin{displaymath}
f(\bxi, I) = \sum_{\alpha \in \bbN^3} \sum_{k \in \bbN}
  f_{\alpha,k}' \psi_{\alpha,k,T_{\mathrm{tr}}',T_{\mathrm{int}}'}
  \left(
    \frac{\bxi - \bu'}{\sqrt{R T_{\mathrm{tr}}'}}, 
    \frac{I^{2/\delta}}{R T_{\mathrm{int}}'}
  \right)
\end{displaymath}
for some given $\bu'$, $T_{\mathrm{tr}}'$ and $T_{\mathrm{int}}'$.
\end{quote}
This can be tackled by two steps. First, we will find a series of
coefficients $f_{\alpha,k}''$, such that
\begin{equation} \label{eq:f'}
f(\bxi, I) = \sum_{\alpha \in \bbN^3} \sum_{k \in \bbN}
  f_{\alpha,k}'' \psi_{\alpha,k,T_{\mathrm{tr}},T_{\mathrm{int}}'}
  \left(
    \frac{\bxi - \bu}{\sqrt{R T_{\mathrm{tr}}}}, 
    \frac{I^{2/\delta}}{R T_{\mathrm{int}}'}
  \right).
\end{equation}
Note that only $f_{\alpha,k}$'s with $k = 0,1$ are interested, which
significantly reduces the difficulty of the problem. We calculate
\begin{equation}
\Phi_{\alpha,k} := \int_{\bbR^3 \times \bbR^+}
  f(\bxi, I) \psi_{\alpha,k,T_{\mathrm{tr}},T_{\mathrm{ref}}} \left(
    \frac{\bxi - \bu}{\sqrt{R T_{\mathrm{tr}}}},
    \frac{I^{2/\delta}}{R T_{\mathrm{ref}}}
  \right) \exp \left(
    \frac{|\bxi - \bu|^2}{2 R T_{\mathrm{tr}}} +
    \frac{I^{2/\delta}}{R T_{\mathrm{ref}}}
  \right) \dd\bxi \dd I
\end{equation}
using both \eqref{eq:expansion} and \eqref{eq:f'}, where
$T_{\mathrm{ref}}$ is an arbitrary positive real number. Thanks to the
orthogonality of Hermite and Laguerre polynomials, the results can be
obtained as
\begin{align}
\label{eq:psi_0}
\Phi_{\alpha,0} & = C_0 f_{\alpha,0} = C_0 f_{\alpha,0}'', \\
\label{eq:psi_1}
\Phi_{\alpha,1} & = C_1 \left[
  f_{\alpha,1} +
  \frac{\delta}{2} (R T_{\mathrm{ref}} - R T_{\mathrm{int}})
    f_{\alpha,0}
\right] = C_1 \left[
  f_{\alpha,1}'' +
  \frac{\delta}{2} (R T_{\mathrm{ref}} - R T_{\mathrm{int}}')
    f_{\alpha,0}''
\right],
\end{align}
where
\begin{align}
C_0 &= \frac{2}{\delta} \frac{\alpha!}{(2\pi)^{3/2}}
  \Gamma(m+1) (R T_{\mathrm{tr}})^{-(|\alpha|+3)}
  (R T_{\mathrm{ref}})^{-\delta/2}, \\
C_1 &= \frac{2}{\delta} \frac{\alpha!}{(2\pi)^{3/2}}
  \Gamma(m+2) (R T_{\mathrm{tr}})^{-(|\alpha|+3)}
  (R T_{\mathrm{ref}})^{-(\delta/2+2)}.
\end{align}
With \eqref{eq:psi_0} and \eqref{eq:psi_1}, we immediately get
\begin{equation} \label{eq:f''}
f_{\alpha,0}'' = f_{\alpha,0}, \qquad
f_{\alpha,1}'' = f_{\alpha,1} + \frac{\delta}{2}
  (R T_{\mathrm{int}}' - R T_{\mathrm{int}}) f_{\alpha,0}.
\end{equation}
The second step is to calculate $f_{\alpha,k}'$ from $f_{\alpha,k}''$.
Since the scaling factor on the $I$-axis $R T_{\mathrm{int}}'$ is not
changed in such transformation, we can use the very technique
introduced in \cite{NRxx} to obtain $f_{\alpha,k}'$ efficiently.
Specifically speaking, we define
\begin{equation}
F(\bv, J, \tau) = \sum_{\alpha \in \bbN^3} \sum_{k \in \bbN}
  F_{\alpha,k}(\tau) [(\hat{T} - 1) \tau + 1]^{|\alpha| + 3}
  \psi_{\alpha,k,T_{\mathrm{tr}},T_{\mathrm{int}}'} \left(
    [(\hat{T} - 1) \tau + 1] \bv + \tau \bw, J
  \right),
\end{equation}
where
\begin{equation}
\hat{T} = \sqrt{\frac{T_{\mathrm{tr}}}{T_{\mathrm{tr}}'}}, \qquad
\bw = \frac{\bu - \bu'}{\sqrt{R T_{\mathrm{tr}}'}}, \qquad
\tau \in [0,1].
\end{equation}
If we require
\begin{equation}
\frac{\partial F}{\partial \tau} \equiv 0,
  \quad \forall \tau \in [0,1], \qquad \text{and} \qquad
F_{\alpha,k}(0) = f_{\alpha,k}'',
  \quad \forall (\alpha,k) \in \bbN^3 \times \bbN,
\end{equation}
then it is easy to find $F_{\alpha,k}(1) = f_{\alpha,k}'$, $\forall
(\alpha,k) \in \bbN^3 \times \bbN$. The next job is to write
$\frac{\partial F}{\partial \tau}$ in the form of
\eqref{eq:expansion}, and then require each coefficient to be zero.
Thus an ordinary differential system is obtained. The calculation of
$\frac{\partial F}{\partial \tau}$ is almost a repetition of that
presented in \cite{NRxx}, which is omitted here. The resulting
ordinary differential equations are
\begin{equation} \label{eq:ode}
\left\{ \begin{array}{l}
\displaystyle \frac{\mathrm{d} F_{\alpha,k}}{\mathrm{d} \tau} =
  [1 - \tau S(\tau)]^2 \sum_{d=1}^3 \left[
    S(\tau) R T_{\mathrm{tr}} F_{\alpha-2e_d,k} +
    (u_d - u_d') \hat{T} F_{\alpha-e_d,k}
  \right], \\
F_{\alpha,k}(0) = f_{\alpha,k}'',
\end{array} \right.
\end{equation}
where
\begin{equation}
S(\tau) = \frac{\hat{T} - 1}{(\hat{T} - 1) \tau + 1}.
\end{equation}
In our implementation, we solve \eqref{eq:ode} using Runge-Kutta
method until $\tau = 1$, and then $F_{\alpha,k}(1) = f_{\alpha,k}'$
can be obtained. The readers can find some properties of
\eqref{eq:ode} in \cite{NRxx}. Equations \eqref{eq:f''} and
\eqref{eq:ode} give a practical way to calculate $f_{\alpha,k}'$ for
$k=0,1$.

\subsubsection{The collision step}
It remains to give a numerical method to solve the collision-only
equation
\begin{equation} \label{eq:collision_only_Boltzmann}
\frac{\partial f}{\partial t} = \frac{1}{\varepsilon} (G - f)
\end{equation}
by one time step. The corresponding moment equations can be extracted
from \eqref{eq:mnt_eqs} by removing the terms arising in the
convection term. This results
\begin{equation} \label{eq:collision_only}
\frac{\partial f_{\alpha,k}}{\partial t}
  + \sum_{d=1}^3 \frac{\partial u_d}{\partial t} f_{\alpha-e_d,k}
  + \frac{1}{2} \frac{\partial (R T_{\mathrm{tr}})}{\partial t}
    \sum_{d=1}^3 f_{\alpha-2e_d,k}
  - (m+k) \frac{\partial (R T_{\mathrm{int}})}{\partial t}
    f_{\alpha, k-1}
= \frac{1}{\varepsilon} (G_{\alpha,k} - f_{\alpha,k}).
\end{equation}
Let $(\alpha, k) = (0,0)$ and $(\alpha, k) = (e_j, 0)$ respectively,
and one has
\begin{equation} \label{eq:col_inv_rho_u}
\frac{\partial \rho}{\partial t} = 0, \qquad
\frac{\partial u_j}{\partial t} = 0.
\end{equation}
Thus the second term in \eqref{eq:collision_only} vanishes. Let
$(\alpha, k) = (2e_j, 0)$, $j = 1,2,3$ and sum the equations up, and
we get
\begin{equation}
\frac{\partial T_{\mathrm{tr}}}{\partial t} =
  \frac{1}{\varepsilon Z} (T_{\mathrm{eq}} - T_{\mathrm{tr}})
\end{equation}
Setting $(\alpha, k) = (0, 1)$ in \eqref{eq:collision_only}, it
becomes
\begin{equation}
\frac{\partial T_{\mathrm{int}}}{\partial t} =
  \frac{1}{\varepsilon Z} (T_{\mathrm{eq}} - T_{\mathrm{int}}).
\end{equation}
Using \eqref{eq:T_eq}, it is easy to obtain
\begin{equation} \label{eq:col_inv_T_eq}
\frac{\partial T_{\mathrm{eq}}}{\partial t} = 0.
\end{equation}
Equations \eqref{eq:col_inv_rho_u} and \eqref{eq:col_inv_T_eq} agree
with the fact that the density, velocity and temperature are not
changed by collision. Now \eqref{eq:collision_only} can be rewritten
as
\begin{equation} \label{eq:collision_only_f}
\frac{\partial f_{\alpha,k}}{\partial t} = \frac{1}{\varepsilon} \left[
  G_{\alpha,k} - f_{\alpha,k} -
  Z^{-1} (R T_{\mathrm{eq}} - R T_{\mathrm{tr}}) \left(
    \frac{1}{2} \sum_{j=1}^3 f_{\alpha - 2e_j,k} -
    (m + k) f_{\alpha, k - 1}
  \right)
\right].
\end{equation}
Setting $(\alpha,k) = (e_i + e_j,0)$ in \eqref{eq:G_recur} and
substituting the result into \eqref{eq:collision_only_f}, one finds
that the collision-only equation for $f_{e_i + e_j,0}$ has also a
simple form:
\begin{equation} \label{eq:collision_only_Theta}
\frac{\partial f_{e_i + e_j,0}}{\partial t} =
  -\frac{1}{\varepsilon \mathrm{Pr}} f_{e_i + e_j,0},
  \quad i,j = 1,2,3,
\end{equation}
which agrees with the settings in \cite{Brull}. Now suppose we want to
solve \eqref{eq:collision_only_Boltzmann} from $t^n$ to $t^{n+1}$. For
simplicity, for any quantity $\psi$, we use $\psi^n$, $\psi^{n+1}$ and
$\psi^{n+1/2}$ to denote $\psi(t^n)$, $\psi(t^{n+1})$ and
$\psi\left(\frac{1}{2} (t^n + t^{n+1}) \right)$, respectively. Then
the following relations holds analytically for $i,j = 1,2,3$ and $t
\in [t^n, t^{n+1}]$:
\begin{equation} \label{eq:ana_sol}
\begin{gathered}
\rho(t) \equiv \rho^n, \qquad
u_j(t) \equiv u_j^n, \qquad 
T_{\mathrm{eq}}(t) \equiv T_{\mathrm{eq}}^n, \\
T_{\mathrm{tr}}(t) = T_{\mathrm{eq}}^n +
  (T_{\mathrm{tr}}^n - T_{\mathrm{eq}}^n) \exp \left(
    -\frac{t - t^n}{\varepsilon Z}
  \right), \\
T_{\mathrm{int}}(t) = T_{\mathrm{eq}}^n +
  (T_{\mathrm{int}}^n - T_{\mathrm{eq}}^n) \exp \left(
    -\frac{t - t^n}{\varepsilon Z}
  \right), \\
f_{e_i + e_j,0}(t) = f_{e_i + e_j,0}^n \exp \left(
  -\frac{t - t^n}{\varepsilon \mathrm{Pr}}
\right).
\end{gathered}
\end{equation}
These are deduced from \eqref{eq:col_inv_rho_u}---\eqref{eq:col_inv_T_eq}
and \eqref{eq:collision_only_Theta}. Based on \eqref{eq:ana_sol},
$G_{\alpha,k}(t)$ can also be obtained since it can be observed from 
\eqref{eq:G_alpha_k} and \eqref{eq:G_alpha_1}---\eqref{eq:G_recur}
that $G_{\alpha,k}$ are fully determined by the variables listed in
\eqref{eq:ana_sol}. For other cases, meaning $|\alpha| > 2$ if $k = 0$
and $|\alpha| > 0$ if $k = 1$, the Crank-Nicolson scheme is employed
to give a numerical approximation of \eqref{eq:collision_only_f}:
\begin{equation} \label{eq:Crank-Nicolson}
\begin{split}
\frac{f_{\alpha,k}^{n+1} - f_{\alpha,k}^n}{\Delta t} &=
\frac{1}{\varepsilon} \Bigg[
  G_{\alpha, k}^{n+1/2} -
  \frac{f_{\alpha,k}^{n+1} + f_{\alpha,k}^n}{2} - 
Z^{-1} \left(
    R T_{\mathrm{eq}}^{n+1/2} -
    R T_{\mathrm{tr}}^{n+1/2}
  \right) \times \\
& \qquad \qquad \Bigg(
  \frac{1}{2} \sum_{j=1}^3
    \frac{f_{\alpha - 2e_j,k}^{n+1} + f_{\alpha - 2e_j,k}^n}{2}
  - (m+k) \frac{f_{\alpha, k-1}^{n+1} + f_{\alpha, k-1}^n}{2}
\Bigg) \Bigg],
\end{split}
\end{equation}
where $\Delta t = t^{n+1} - t^n$. Note that no linear system needs to
be solved when applying \eqref{eq:Crank-Nicolson}, since when solving
$f_{\alpha,k}^{n+1}$, the terms $f_{\alpha-2e_j, k}^{n+1}$ and
$f_{\alpha,k-1}^{n+1}$ have always been obtained, and then
\eqref{eq:Crank-Nicolson} is simply a linear equation of
$f_{\alpha,k}^{n+1}$.

\begin{remark}
When $\mathrm{Pr} = Z = 1$, one has $\mathcal{T} = R T_{\mathrm{eq}}
\mathrm{Id}$ and $T_{\mathrm{rel}} = T_{\mathrm{eq}}$ in \eqref{eq:G}.
Obviously, in this case, the ES-BGK model reduces to the BGK model,
and the above technique for processing the collision terms is still
valid.
\end{remark}


\section{Numerical examples} \label{sec:num}
In this section, two one-dimensional numerical examples are presented to
validate our algorithm. The spatial variable $\bx$ will be written in
plain font as $x$. For both tests, the non-dimensional form of the
Boltzmann equation \eqref{eq:ES-BGK} is used. Thus the gas constant
$R$ is chosen as $1$, and the Knudsen number $\Kn$, which is the ratio
of the mean free path to the characteristic length, controls the
rarefaction of the gas. Only the diatomic gas is considered below;
therefore $\delta$ equals to $2.0$. The CFL number is set to be $0.95$
in all runs.

\subsection{Shock tube test}
There have been a number of studies on using the moment method to
solve shock tube problems. In \cite{Torrilhon2000}, the 13-moment case
is carefully investigated and the numerical results in \cite{Weiss,
Au} indicate that the theory for 13-moment case can also be applied
to systems with almost any number of moments. And in \cite{NRxx,
NRxx_new}, the shock tube problem in the monatomic case is calculated
to show the convergence of the original {\NRxx} method and its
improved version when the number of moments increases. Here the
similar settings are used the test the polyatomic {\NRxx} method. The
initial conditions are
\begin{equation}
f(0, x, \bxi, I) = \left\{ \begin{array}{ll}
  \rho_l \psi_{0,0,T_l,T_l} ( \bxi/\sqrt{T_l}, I/T_l ), & x < 0, \\
  \rho_r \psi_{0,0,T_r,T_r} ( \bxi/\sqrt{T_r}, I/T_r ), & x > 0,
\end{array} \right.
\end{equation}
where $\rho_l = 7$, $\rho_r = 1$, and $T_l = T_r = 1$. Recalling that
$R = 1$ and $\delta = 2$, we find the fluid states on both left and
right sides are in equilibrium. As an artificial test, a simple
expression for the viscosity coefficient $\mu$ is chosen as
\begin{equation}
\mu = \Kn \cdot T_{\mathrm{eq}}.
\end{equation}
Four additional parameters, including the Prandtl number
$\mathrm{Pr}$, the relaxation collision number $Z$, the Knudsen number
$\Kn$, and the maximal moment order $M_0$, need to be defined for this
problem. Below, different combinations of these parameters are tested
in our numerical examples to show different properties of the
polyatomic {\NRxx} method. In all the experiments, the computational
domain is $[-2, 2]$ and discretized using $400$ uniform spatial grids.
In the following subsections, all plots show the numerical results at
$t = 0.3$.

\subsubsection{Convergence in the number of moments}
In this part, we set $\mathrm{Pr} = 0.72$ and $Z = 5$, and test the
behavior of solutions for several Knudsen numbers when $M_0$
increases.  In order to provide a reference solution, Mieussens'
conservative discrete velocity model (CDVM) \cite{Mieussens, Dubroca}
is computed.  For CDVM, we use the technique of dimension reduction to
speed up the computation. That is, we define
\begin{equation}
\begin{aligned}
g(t, x, \xi_1) &= \int_{\bbR^2 \times \bbR^+}
  f(t, x, \bxi, I) \dd \xi_2 \dd \xi_3 \dd I, \\
h_1(t, x, \xi_1) &= \frac{1}{2} \int_{\bbR^2 \times \bbR^+}
  (\xi_2^2 + \xi_3^2) f(t, x, \bxi, I) \dd \xi_2 \dd \xi_3 \dd I, \\
h_2(t, x, \xi_1) &= \int_{\bbR^2 \times \bbR^+}
  I f(t, x, \bxi, I) \dd \xi_2 \dd \xi_3 \dd I,
\end{aligned}
\end{equation}
and solve $g$, $h_1$ and $h_2$ instead of the distributions with full
three-dimensional microscopic velocity. The velocity space is
discretized by $400$ grids. Currently, this skill is not used in the
{\NRxx} method.

First, a relatively dense case $\Kn = 0.05$ is considered. The density
and equilibrium temperature results for $M_0 = 3,\cdots,8$ are given
in Figure \ref{fig:Kn=0.05}. For $M_0 \geqslant 5$, the {\NRxx}
results match with the CDVM results very well. This agrees with the
observation in \cite{NRxx, NRxx_new} that for small Knudsen numbers, a
small number of moments can describe the macroscopic quantities in a
high accuracy. The results for a rarefied case $\Kn = 0.5$ from
$M_0=3$ to $M_0=20$ are presented in Figure \ref{fig:Kn=0.5}. It is
obvious that in order to match the CDVM results, much greater number of
moments are needed. However, we can still find the {\NRxx} results
converge as $M_0$ increases, and the limit is probably the solution of
the Boltzmann equation.

\newcommand{\figrT}[1]{
\begin{overpic}[width=.47\textwidth]{#1}
\put(6,40){$\rho$}
\put(94,40){$T_{\mathrm{eq}}$}
\put(50.5,1){$x$}
\end{overpic}
}

\begin{figure}[!ht]
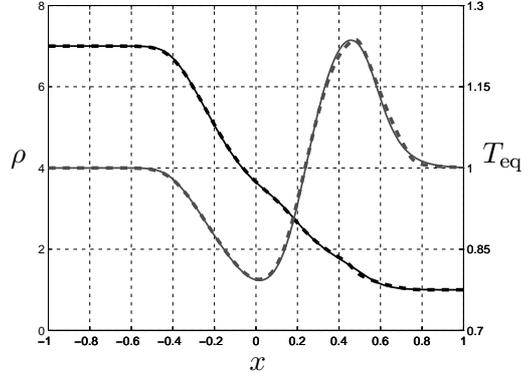
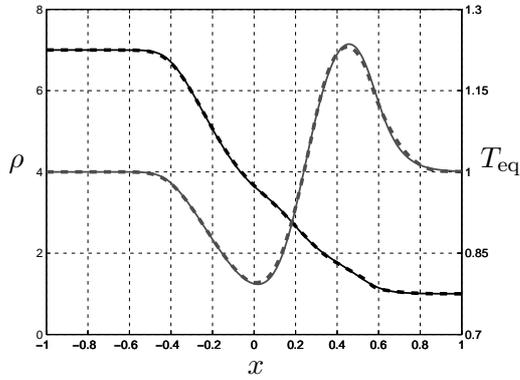
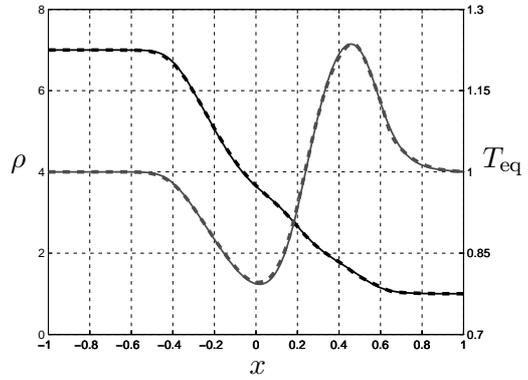
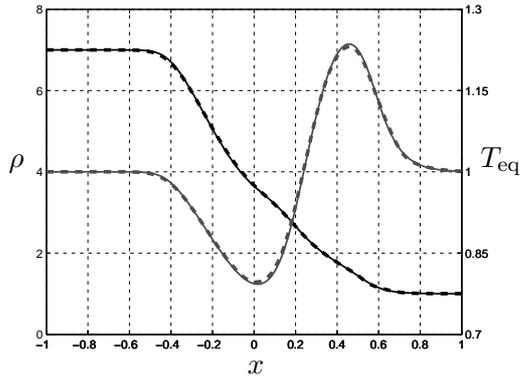
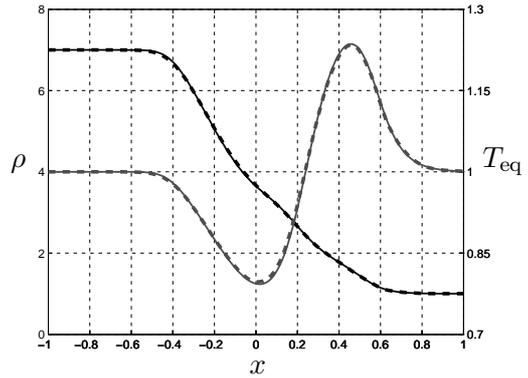

\centering
\subfigure[$M_0=3$, $24$ moments]{ \figrT{Kn=0.05_M=3.eps} }%
\subfigure[$M_0=4$, $45$ moments]{ \figrT{Kn=0.05_M=4.eps} }\\
\subfigure[$M_0=5$, $76$ moments]{ \figrT{Kn=0.05_M=5.eps} }%
\subfigure[$M_0=6$, $119$ moments]{ \figrT{Kn=0.05_M=6.eps} }\\
\subfigure[$M_0=7$, $176$ moments]{ \figrT{Kn=0.05_M=7.eps} }%
\subfigure[$M_0=8$, $249$ moments]{ \figrT{Kn=0.05_M=8.eps} }
\caption{Numerical results for the shock tube problem with $\Kn =
0.05$. The dashed lines are the {\NRxx} results, and the solid thin
lines are the CDVM results with linearization. The dashdot lines are
the results of discrete velocity model. The black lines denote the
density $\rho$ and the gray lines denote the equilibrium temperature
$T_{\mathrm{eq}}$.}
\label{fig:Kn=0.05}
\end{figure}

\begin{figure}[!ht]
\centering
\subfigure[$M_0=3$, $24$ moments]{ \figrT{Kn=0.5_M=3.eps} }%
\subfigure[$M_0=4$, $45$ moments]{ \figrT{Kn=0.5_M=4.eps} }\\
\subfigure[$M_0=5$, $76$ moments]{ \figrT{Kn=0.5_M=5.eps} }%
\subfigure[$M_0=6$, $119$ moments]{ \figrT{Kn=0.5_M=6.eps} }\\
\subfigure[$M_0=7$, $176$ moments]{ \figrT{Kn=0.5_M=7.eps} }%
\subfigure[$M_0=8$, $249$ moments]{ \figrT{Kn=0.5_M=8.eps} }
\caption{Numerical results for the shock tube problem with $\Kn =
0.5$. The dashed lines are the {\NRxx} results, and the solid thin
lines are the CDVM results with linearization. The dashdot lines are
the results of discrete velocity model. The black lines denote the
density $\rho$ and the gray lines denote the equilibrium temperature
$T_{\mathrm{eq}}$ (continued on the next page).}
\label{fig:Kn=0.5}
\end{figure}

\addtocounter{figure}{-1}
\begin{figure}[!ht]
\let\oldcaptionlabeldelim\captionlabeldelim
\renewcommand\captionlabeldelim{~(continued)\oldcaptionlabeldelim}
\addtocounter{subfigure}{6}
\centering
\subfigure[$M_0=9$, $340$ moments]{ \figrT{Kn=0.5_M=9.eps} }%
\subfigure[$M_0=10$, $451$ moments]{ \figrT{Kn=0.5_M=10.eps} }\\
\subfigure[$M_0=11$, $584$ moments]{ \figrT{Kn=0.5_M=11.eps} }%
\subfigure[$M_0=12$, $741$ moments]{ \figrT{Kn=0.5_M=12.eps} }\\
\subfigure[$M_0=13$, $924$ moments]{ \figrT{Kn=0.5_M=13.eps} }%
\subfigure[$M_0=14$, $1135$ moments]{ \figrT{Kn=0.5_M=14.eps} }
\caption{Numerical results for the shock tube problem with $\Kn =
0.5$. The dashed lines are the {\NRxx} results, and the solid thin
lines are the CDVM results with linearization. The dashdot lines are
the results of discrete velocity model. The black lines denote the
density $\rho$ and the gray lines denote the equilibrium temperature
$T_{\mathrm{eq}}$ (continued on the next page).}
\end{figure}

\addtocounter{figure}{-1}
\begin{figure}[!ht]
\let\oldcaptionlabeldelim\captionlabeldelim
\renewcommand\captionlabeldelim{~(continued)\oldcaptionlabeldelim}
\addtocounter{subfigure}{12}
\centering
\subfigure[$M_0=15$, $1376$ moments]{ \figrT{Kn=0.5_M=15.eps} }%
\subfigure[$M_0=16$, $1649$ moments]{ \figrT{Kn=0.5_M=16.eps} }\\
\subfigure[$M_0=17$, $1956$ moments]{ \figrT{Kn=0.5_M=17.eps} }%
\subfigure[$M_0=18$, $2299$ moments]{ \figrT{Kn=0.5_M=18.eps} }\\
\subfigure[$M_0=19$, $2680$ moments]{ \figrT{Kn=0.5_M=19.eps} }%
\subfigure[$M_0=20$, $3101$ moments]{ \figrT{Kn=0.5_M=20.eps} }
\caption{Numerical results for the shock tube problem with $\Kn =
0.5$. The dashed lines are the {\NRxx} results, and the solid thin
lines are the CDVM results with linearization. The dashdot lines are
the results of discrete velocity model. The black lines denote the
density $\rho$ and the gray lines denote the equilibrium temperature
$T_{\mathrm{eq}}$.}
\end{figure}

Now we consider a severe case $\Kn = 5$, and the results are given in
Figure \ref{fig:Kn=5.0}. Although the {\NRxx} solutions deviate from
the reference solution greater than those in Figure \ref{fig:Kn=0.5},
the convergence is again very clear. From Figure \ref{fig:Kn=0.05}---%
\ref{fig:Kn=5.0}, we can find the theory in \cite{Torrilhon2000} is
also valid for the regularized moment methods. For a fixed choice of
$M_0$, the corresponding moment system always fails to describe the
physical phenomenon when $t \rightarrow 0$ (or $\Kn \rightarrow
\infty$ for a fixed time $t$) due to the very strong non-equilibrium.
As $t$ increases, the collision term starts to show an effect of
dissipation, and the solution of the moment system gradually presents
its physical meaning. As is shown in \cite{Weiss, Au}, for a greater
$M_0$, such progress is faster. Our numerical results correctly
exhibit this behavior.

\begin{figure}[!ht]
\centering
\subfigure[$M_0=3$, $24$ moments]{ \figrT{Kn=5.0_M=3.eps} }%
\subfigure[$M_0=6$, $119$ moments]{ \figrT{Kn=5.0_M=6.eps} }\\
\subfigure[$M_0=9$, $340$ moments]{ \figrT{Kn=5.0_M=9.eps} }%
\subfigure[$M_0=12$, $741$ moments]{ \figrT{Kn=5.0_M=12.eps} }\\
\caption{Numerical results for the shock tube problem with $\Kn =
5.0$. The dashed lines are the {\NRxx} results, and the solid thin
lines are the CDVM results with linearization. The dashdot lines are
the results of discrete velocity model. The black lines denote the
density $\rho$ and the gray lines denote the equilibrium temperature
$T_{\mathrm{eq}}$ (continued on the next page).}
\label{fig:Kn=5.0}
\end{figure}

\addtocounter{figure}{-1}
\begin{figure}[!ht]
\let\oldcaptionlabeldelim\captionlabeldelim
\renewcommand\captionlabeldelim{~(continued)\oldcaptionlabeldelim}
\addtocounter{subfigure}{4}
\centering
\subfigure[$M_0=15$, $1376$ moments]{ \figrT{Kn=5.0_M=15.eps} }%
\subfigure[$M_0=18$, $2299$ moments]{ \figrT{Kn=5.0_M=18.eps} }\\
\subfigure[$M_0=21$, $3564$ moments]{ \figrT{Kn=5.0_M=21.eps} }%
\subfigure[$M_0=24$, $5225$ moments]{ \figrT{Kn=5.0_M=24.eps} }
\caption{Numerical results for the shock tube problem with $\Kn =
5.0$. The dashed lines are the {\NRxx} results, and the solid thin
lines are the CDVM results with linearization. The dashdot lines are
the results of discrete velocity model. The black lines denote the
density $\rho$ and the gray lines denote the equilibrium temperature
$T_{\mathrm{eq}}$.}
\end{figure}

\subsubsection{Comparison between BGK and ES-BGK collision terms}
As is known, for monatomic gases, the BGK model fails to predict the
correct Prandtl number, while $\mathrm{Pr}$ is considered as a
parameter in the ES-BGK collision term. For polyatomic gases, besides
the Prandtl number, the BGK model also gives incorrect relaxation
collision number $Z$. Actually, the BGK model always gives $Z = 1$,
which means the translational and internal temperatures tend to the
equilibrium temperature more rapidly than the ES-BGK model. Thus it
can be expected that the BGK model gives incorrect translational
temperature, internal temperature, and heat fluxes.

\begin{figure}[!ht]
\centering
\begin{overpic}[width=.7\textwidth]{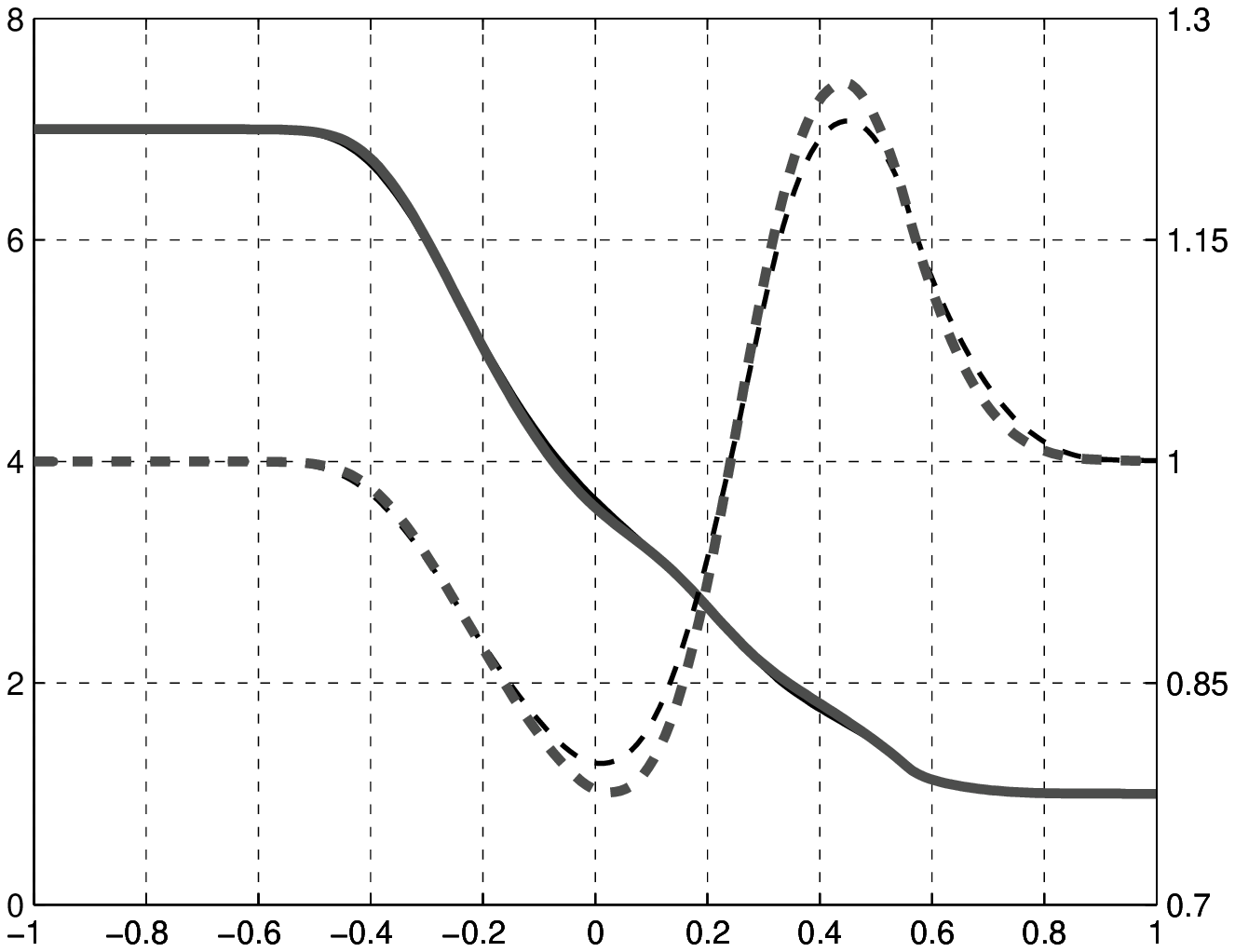}
\put(6,40){$\rho$}
\put(94,40){$T_{\mathrm{eq}}$}
\put(50.5,1){$x$}
\end{overpic}
\caption{Comparison between BGK and ES-BGK models. The black lines are
the results of the ES-BGK model, and the gray lines are the results of
the BGK model. The solid lines denote the profile of density $\rho$,
and the dashed lines denote the profile of equilibrium temperature
$T_{\mathrm{eq}}$.}
\label{fig:rho_T}
\end{figure}

\begin{figure}[!ht]
\centering
\begin{overpic}[width=.7\textwidth]{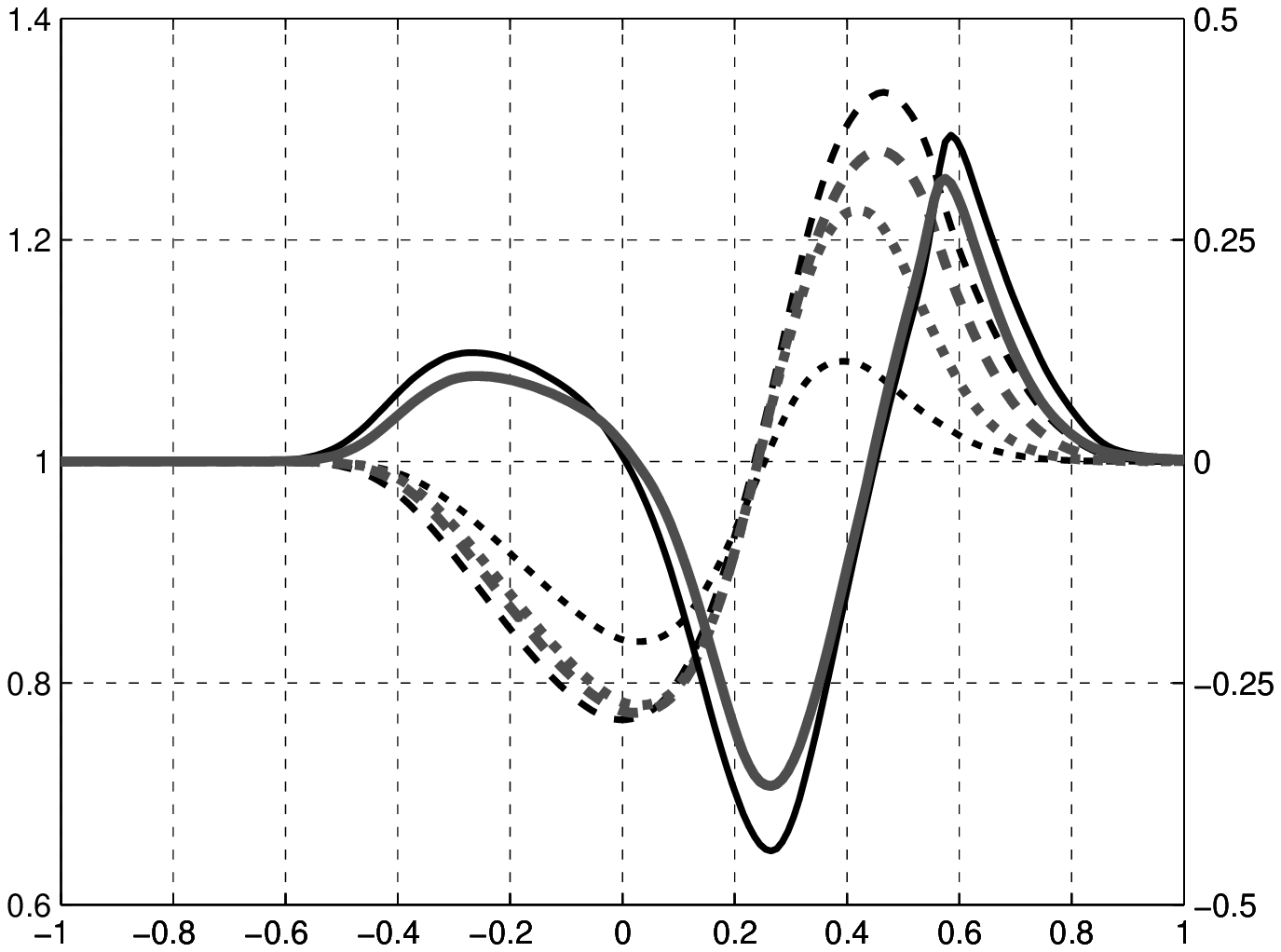}
\put(-2,40){$T_{\mathrm{tr}}, T_{\mathrm{int}}$}
\put(95,40){$q_1$}
\put(50.5,1){$x$}
\end{overpic}
\caption{Comparison between BGK and ES-BGK models. The black lines are
the results of the ES-BGK model, and the gray lines are the results of
the BGK model. The dashed and dotted lines denote the profile of
translational temperature $T_{\mathrm{tr}}$ and internal temperature
$T_{\mathrm{int}}$, respectively, and the solid lines denote the heat
flux $q_1$.}
\label{fig:T_q}
\end{figure}

As a test, we set $\mathrm{Pr} = 0.72$, $Z = 5$, $\Kn = 0.05$ and $M_0
= 5$, and both BGK and ES-BGK collision models are computed. The
results are shown in Figure \ref{fig:rho_T} and Figure \ref{fig:T_q}.
In Figure \ref{fig:rho_T}, it is found that the BGK model gives a
pretty good prediction of the density, whereas the deviation of
temperature between two models is significant. Figure \ref{fig:T_q}
shows that the BGK model provides much smaller difference between the
translational temperature and the internal temperature than the
ES-BGK model, which indicates different relaxation collision numbers
involved in the two models. The heat flux $q_1$ is defined as
\begin{equation}
q_1 = \int_{\bbR^3 \times \bbR^+} (\xi_1 - u_1) \left(
  \frac{1}{2} |\bxi - \bu|^2 + I^{2/\delta}
\right) f(\bxi, I) \dd \bxi \dd I.
\end{equation}
The difference in the heat flux is caused by the discordance of both
the Prandtl number and the relaxation collision number.

\subsubsection{Comparison between the monatomic case and the polyatomic case}
Let $Z = \infty$ and define
\begin{equation}
g(t, \bx, \bxi) = \int_{\bbR^+} f(t, \bx, \bxi, I) \dd I.
\end{equation}
Integrating the both sides of \eqref{eq:ES-BGK} over $\bbR^+$ with
respect to $I$, it is not difficult to find that the reduced
distribution function $g$ satisfies the monatomic Boltzmann equation
with ES-BGK collision operator. Thus, it is natural to expect that
when the relaxation collision number $Z$ gets greater, the polyatomic
case will get closer to the monatomic case. The part is devoted to the
numerical validation of this behavior.

\begin{figure}[!ht]
\centering
\subfigure[Density profile]{
\includegraphics[width=.7\textwidth]{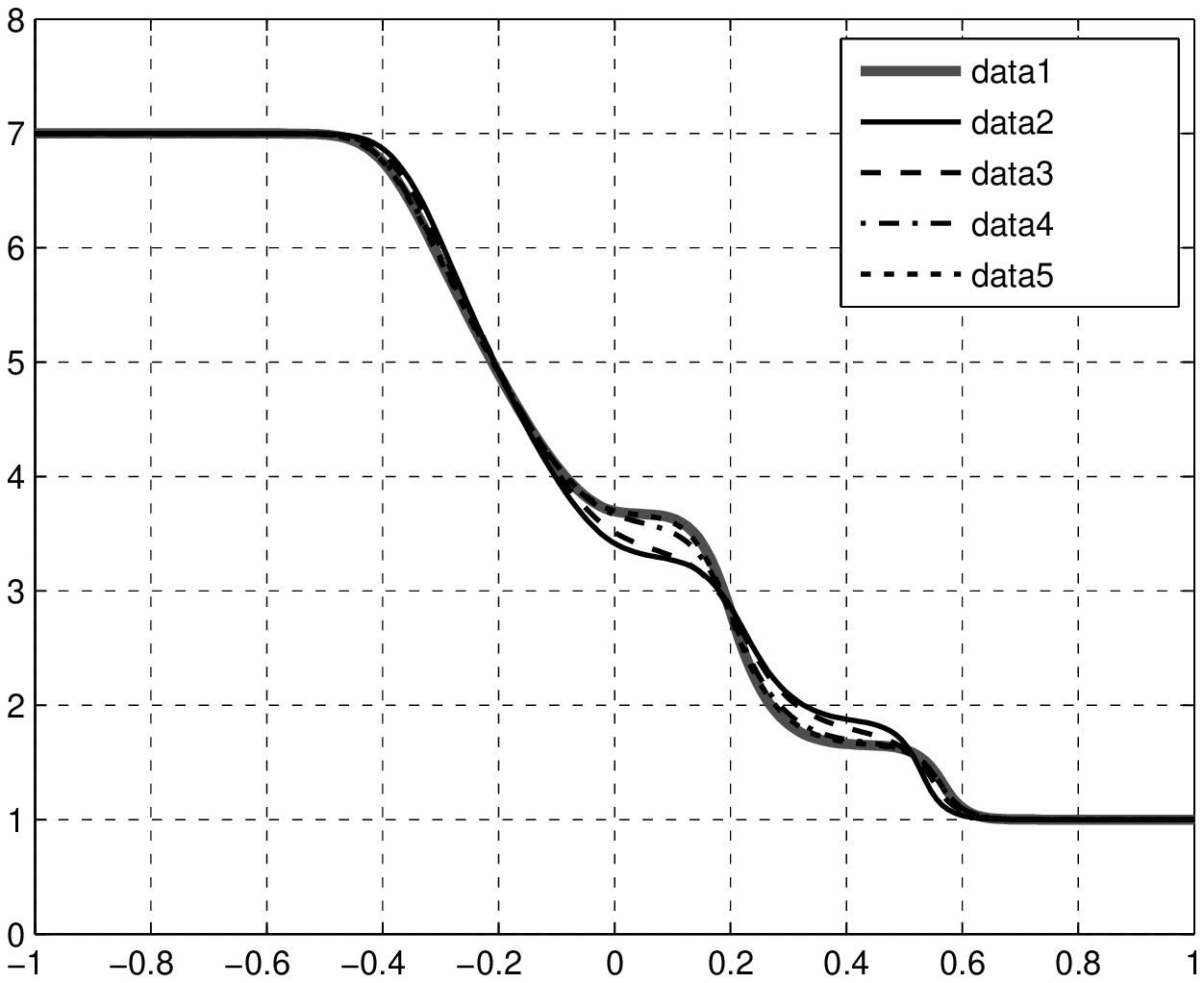}
}
\subfigure[Translational temperature profile]{
\includegraphics[width=.7\textwidth]{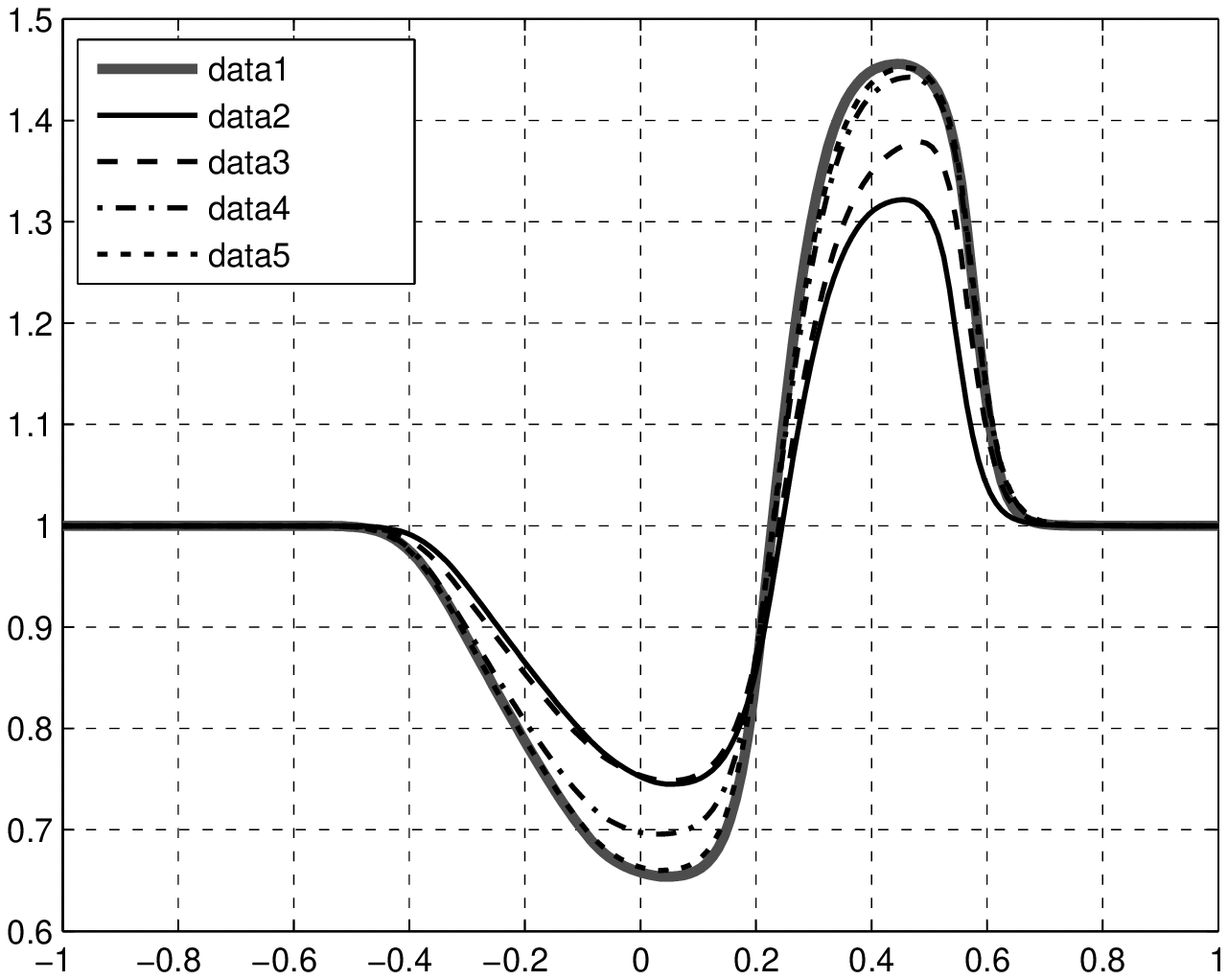}
}
\caption{Comparison between monatomic and polyatomic cases.}
\label{fig:Kn=0.01}
\end{figure}

The {\NRxx} method for monatomic gases has been introduced in
\cite{NRxx, NRxx_new, Cai}, where a BGK collision model is considered.
For the monatomic ES-BGK model, the collision only equation can be
analytically solved, and the result will be reported elsewhere. Four
relaxation collision numbers $Z = 1, 10, 100, 1000$ are considered
here, and other parameters are $\mathrm{Pr} = 2/3$, $\Kn = 0.01$, $M_0
= 5$. The numerical results can be found in Figure \ref{fig:Kn=0.01}.
It clearly shows that the polyatomic result tends to the monatomic
result gradually as $Z$ increases.

\psfrag{data1}{\tiny{monatomic}}
\psfrag{data2}{\tiny{$Z=1$}}
\psfrag{data3}{\tiny{$Z=10$}}
\psfrag{data4}{\tiny{$Z=100$}}
\psfrag{data5}{\tiny{$Z=1000$}}

\subsection{Shock structure of nitrogen}
In this section, we will use the polyatomic {\NRxx} method to
compute the shock structure of nitrogen, trying to reproduce the
experimental results reported in \cite{Alsmeyer}. In order to get a
steady shock structure with Mach number $\Ma$, we solve a Riemann
problem with the following initial condition until a steady state:
\begin{equation}
f(0, x, \bxi, I) = \left\{ \begin{array}{ll}
  \rho_l \psi_{0,0,T_l,T_l} \left(
    (\bxi - \bu_l) / \sqrt{T_l}, I/T_l
  \right), & x < 0, \\
  \rho_r \psi_{0,0,T_r,T_r} \left(
    (\bxi - \bu_r) / \sqrt{T_r}, I/T_r
  \right), & x > 0,
\end{array} \right.
\end{equation}
where
\begin{equation}
\begin{gathered}
\rho_l = 1, \quad \bu_l = (\sqrt{\gamma} \Ma, 0, 0)^T, \quad T_l = 1, \\
\rho_r = \frac{(\gamma + 1) \Ma^2}{(\gamma - 1) \Ma^2 + 2}, \quad
\bu_r = \frac{\rho_l}{\rho_r} \bu_l, \quad
T_r = \frac{2\gamma \Ma^2 - (\gamma - 1)}{(\gamma + 1) \rho_r}.
\end{gathered}
\end{equation}
Here $\gamma$ is the adiabatic index. For nitrogen, $\gamma$ equals to
$1.4$. The Prandtl number is chosen as $0.72$ as in \cite{Tang}. The
Knudsen number is $\Kn = 0.1$, and the grid size is $\Delta x =
0.005$. The computational domain is $[-1.5, 1.5]$, which is large
enough to cover the whole shock structure.

It remains to give the expressions of $Z$ and $\mu$. They have
significant influence on the thickness of the shock. For the
relaxation collision number $Z$, both the gas-kinetic model
\cite{Tang} and the direct simulation Monte Carlo (DSMC)
\cite{Alsmeyer} show that $Z = 4$ or $Z = 5$ best fits the
experimental data. However, in both \cite{Larina} and \cite{Bourgat},
where the Rykov and ES-BGK models are used respectively, it is
reported that a smaller $Z$ between $2$ and $3$ gives better numerical
results.  The same conclusion is drawn by our numerical experiments.
Since \cite{Larina} also considers the shock structure problem, we use
the same settings here:
\begin{equation}
\mu = \frac{5}{8} \sqrt{\frac{\pi}{2}} \Kn T_{\mathrm{eq}}^{0.72},
\qquad Z = 1.45 \left(
  1 + 0.75 \frac{T_{\mathrm{int}}}{T_{\mathrm{tr}}}
\right).
\end{equation}

Six Mach numbers ranging from $\Ma = 1.53$ to $\Ma = 6.1$ are taken
into account. Similar as \cite{NRxx_new}, in order to avoid the
problem of hyperbolicity, only the $24$ moment system ($M_0=3$) is
used in our computation. The numerical results are plotted in Figure
\ref{fig:shock}, where all macroscopic variables are normalized so
that the computational results can match the data in \cite{Alsmeyer}.
Precisely, we use
\begin{equation}
\hat{\rho} = \frac{\rho - \rho_l}{\rho_r - \rho_l}, \quad
\hat{T}_{\mathrm{tr}} = \frac{T_{\mathrm{tr}} - T_l}{T_r - T_l}, \quad
\hat{T}_{\mathrm{int}} = \frac{T_{\mathrm{int}} - T_l}{T_r - T_l},
\end{equation}
and $\lambda$ denotes the mean free path. It can be found that the
density profiles are in very good agreements with the experimental
data and no subshocks exist in the shock structure. With increasing
Mach number, the numerical result gradually deviates from the
experimental data. Only when $\Ma$ is as great as $6.1$, the deviation
in the low density region (around $x/\lambda \in (-4, -1)$ in the
figure) is becoming obvious.

\psfrag{data1}{\scalebox{0.6}{~$\hat{T}_{\mathrm{tr}}$}}
\psfrag{data2}{\scalebox{0.6}{~$\hat{T}_{\mathrm{int}}$}}
\psfrag{data3}{\scalebox{0.6}{~$\hat{\rho}$}}
\psfrag{data4}{\scalebox{0.6}{~$\hat{\rho}$, exp. data}}

\newcommand{\figshock}[1]{
\begin{overpic}[width=.47\textwidth]{#1}
\put(48,0){\small $x / \lambda$}
\end{overpic}
}

\begin{figure}[!ht]
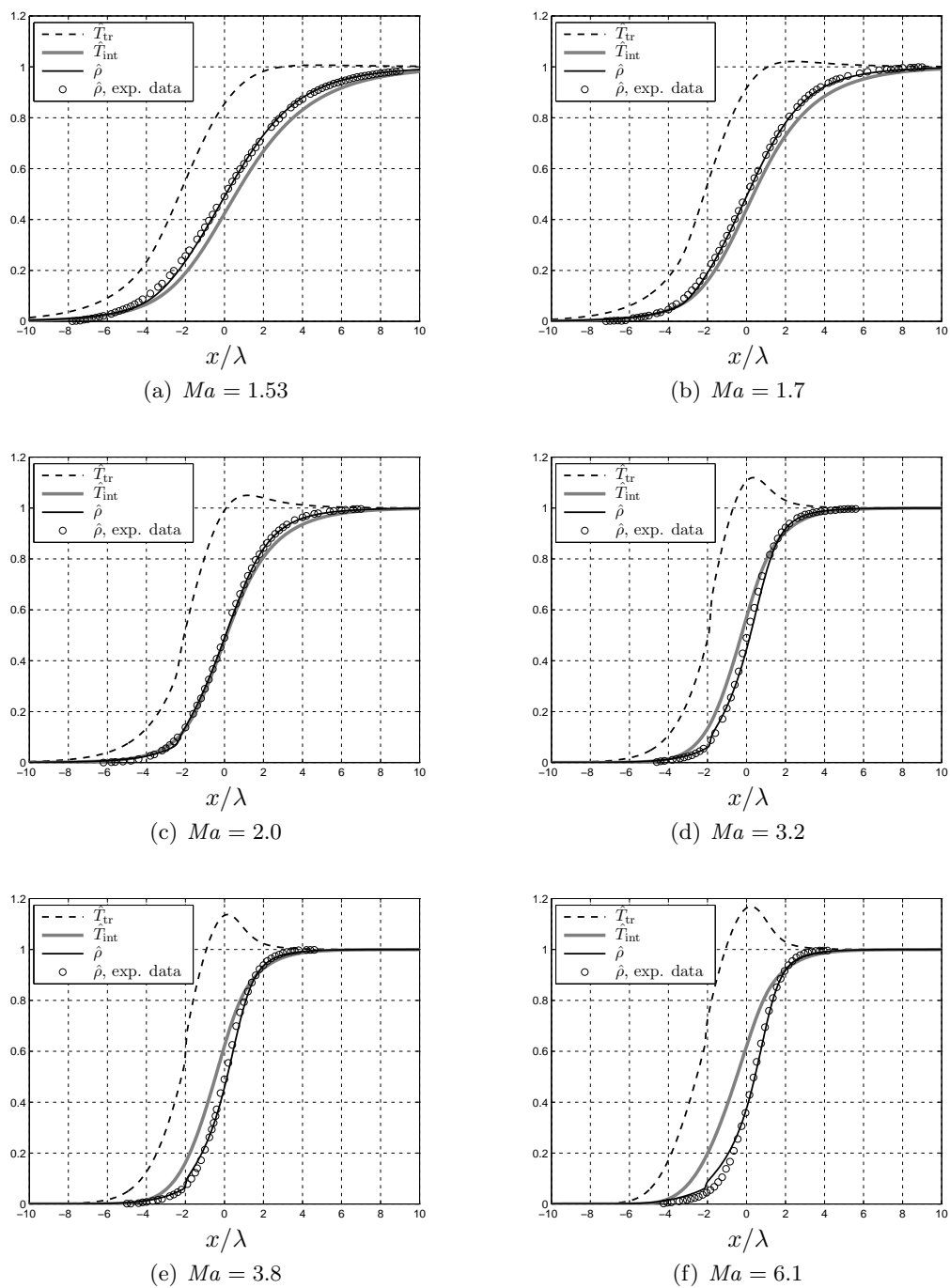

\centering
\subfigure[$\Ma=1.53$]{\figshock{M_1.53.eps}}%
\subfigure[$\Ma=1.7$]{\figshock{M_1.7.eps}} \\
\subfigure[$\Ma=2.0$]{\figshock{M_2.0.eps}}%
\subfigure[$\Ma=3.2$]{\figshock{M_3.2.eps}}\\
\subfigure[$\Ma=3.8$]{\figshock{M_3.8.eps}}%
\subfigure[$\Ma=6.1$]{\figshock{M_6.1.eps}}
\caption{Structure of the nitrogen shock wave. All quantities are
normalized.}
\label{fig:shock}
\end{figure}


\section{Concluding remarks} \label{sec:conclusion}
In this paper, the {\NRxx} method is extended to the polyatomic gases.
Further investigations such as the boundary conditions and the
multidimensional simulations are in progress.

\section*{Acknowledgements}
This research was supported in part by the National Basic Research
Program of China (2011CB309704) and Fok Ying Tong Education and NCET
in China.


\appendix
\section{Properties of Hermite and Laguerre polynomials} \label{sec:orth}
The Hermite polynomials defined in \eqref{eq:Hermite} are a set of
orthogonal polynomials over the domain $(-\infty, +\infty)$.
Below we list some of their properties which are used in this paper:
\begin{enumerate}
\item Orthogonality: $\int_{\bbR} \He_{n_1}(x) \He_{n_2}(x)
  \exp(-x^2/2) \dd x = n_1! \sqrt{2\pi} \delta_{n_1 n_2}$;
\item Recursion relation: $\He_{n+1}(x) = x \He_n(x) - n \He_{n-1}(x)$;
\item Differential relation: $\He_n'(x) = n \He_{n-1}(x)$.
\end{enumerate}
All these properties can be found in many mathematical handbooks such
as \cite{Abramowitz}. And the following equality can be derived from
the last two relations:
\begin{equation}
[\He_n(x) \exp(-x^2/2)]' = -\He_{n+1}(x) \exp(-x^2/2).
\end{equation}

As introduced in Section \ref{sec:spectral}, the Laguerre polynomials
defined in \eqref{eq:Laguerre} are orthogonal over $[0, +\infty)$. The
Laguerre polynomials are closely related to the Hermite polynomials,
and they have very similar properties:
\begin{enumerate}
\item Orthogonality: $\int_{\bbR^+} L_{k_1}^{(m)}(x) L_{k_2}^{(m)}(x)
  x^m \exp(-x) \dd x = \gamma_{k_1}^{(m)} \delta_{k_1 k_2}$;
\item Recursion relation: $(k+1) L_{k+1}^{(m)}(x) = (m+1+k-x)
  L_k^{(m)}(x) - x L_{k-1}^{(m+1)}(x)$;
\item Differential relation: $[L_k^{(m)}(x)]' = -L_{k-1}^{(m+1)}(x)$.
\end{enumerate}
And the last two relations give
\begin{equation} \label{eq:Laguerre_diff}
[L_k^{(m)}(x) \exp(-x)]' =
  x^{-1} [(k+1) L_{k+1}^{(m)}(x) - (m+1+k) L_k^{(m)}(x)] \exp(-x).
\end{equation}

\section{The deduction of polyatomic moment equations} \label{sec:mnt_eqs}
In this appendix, we are going to give the detailed deduction of
\eqref{eq:mnt_eqs}. For simplicity, we define
\begin{equation} \label{eq:psi}
\begin{gathered}
\psi_{1, \alpha, T_{\mathrm{tr}}}(\bv) =
  \left( \sqrt{2 \pi} \right)^{-3}
  (R T_{\mathrm{tr}})^{-\frac{|\alpha|+3}{2}}
  \prod_{d=1}^3 \He_{\alpha_d}(v_d)
    \exp \left( -\frac{v_d^2}{2} \right), \\
\psi_{2, k, T_{\mathrm{int}}}(J) =
  \frac{2}{\delta} \left( \gamma_k^{(m)} \right)^{-1}
  (R T_{\mathrm{int}})^{-(\delta/2+k)} L_k^{(m)}(J) \exp(-J).
\end{gathered}
\end{equation}
Thus $\psi_{\alpha,k,T_{\mathrm{tr}}, T_{\mathrm{int}}}(\bv, J) =
\psi_{1,\alpha,T_{\mathrm{tr}}}(\bv) \psi_{2,k,T_{\mathrm{int}}}(J)$.
It has been deduced in \cite{Cai} that
\begin{equation}
\frac{\partial}{\partial \eta} \psi_{1,\alpha,T_{\mathrm{tr}}}
  \left( \frac{\bxi - \bu}{\sqrt{R T_{\mathrm{tr}}}} \right)
= \sum_{d=1}^3 \left[
  \frac{\partial u_d}{\partial \eta}
    \psi_{1,\alpha + e_d,T_{\mathrm{tr}}}
    \left( \frac{\bxi - \bu}{\sqrt{R T_{\mathrm{tr}}}} \right) +
  \frac{1}{2} \frac{\partial (R T_{\mathrm{tr}})}{\partial \eta}
    \psi_{1,\alpha + 2e_d,T_{\mathrm{tr}}}
    \left( \frac{\bxi - \bu}{\sqrt{R T_{\mathrm{tr}}}} \right)
\right],
\end{equation}
where $\eta$ stands for $t$ or $x_j$, $j = 1,2,3$. Now using
\eqref{eq:Laguerre_diff}, we have
\begin{equation} \label{eq:psi_2_diff}
\begin{split}
& \frac{\partial}{\partial \eta} \psi_{2,k,T_{\mathrm{int}}}
  \left( \frac{I^{2/\delta}}{R T_{\mathrm{int}}} \right) \\
{=} & -\frac{2}{\delta} \left( \gamma_k^{(m)} \right)^{-1}
  \left( \frac{\delta}{2} + k \right)
  (R T_{\mathrm{int}})^{-(\delta/2 + k + 1)}
  \frac{\partial (R T_{\mathrm{int}})}{\partial \eta}
  L_k^{(m)} \left( \frac{I^{2/\delta}}{R T_{\mathrm{int}}} \right)
  \exp \left( -\frac{I^{2/\delta}}{R T_{\mathrm{int}}} \right) \\
& + \frac{2}{\delta} \left( \gamma_k^{(m)} \right)^{-1}
  (R T_{\mathrm{int}})^{-(\delta / 2 + k)}
  \frac{R T_{\mathrm{int}}}{I^{2/\delta}}
  \frac{\partial}{\partial \eta} \left(
    \frac{I^{2/\delta}}{R T_{\mathrm{int}}}
  \right)
  \exp \left( -\frac{I^{2/\delta}}{R T_{\mathrm{int}}} \right)
  \times \\
& \qquad \left[
  (k+1) L_{k+1}^{(m)} \left(
    \frac{I^{2/\delta}}{R T_{\mathrm{int}}}
  \right) - (m + 1 + k) L_{k}^{(m)} \left(
    \frac{I^{2/\delta}}{R T_{\mathrm{int}}}
  \right)
\right].
\end{split}
\end{equation}
Since $m = \delta / 2 - 1$ and
\begin{equation}
\frac{R T_{\mathrm{int}}}{I^{2/\delta}}
  \frac{\partial}{\partial \eta} \left(
    \frac{I^{2/\delta}}{R T_{\mathrm{int}}}
  \right)
= -\frac{1}{R T_{\mathrm{int}}}
  \frac{\partial (R T_{\mathrm{int}})}{\partial \eta},
\end{equation}
the equation \eqref{eq:psi_2_diff} can be simplified as
\begin{equation}
\begin{split}
& \frac{\partial}{\partial \eta} \psi_{2,k,T_{\mathrm{int}}}
  \left( \frac{I^{2/\delta}}{R T_{\mathrm{int}}} \right) \\
{=} & -\frac{2}{\delta} \left( \gamma_k^{(m)} \right)^{-1}
  (k + 1) (R T_{\mathrm{int}})^{-(\delta/2 + k + 1)}
  \frac{\partial (R T_{\mathrm{int}})}{\partial \eta}
  L_{k+1}^{(m)} \left( \frac{I^{2/\delta}}{R T_{\mathrm{int}}} \right)
  \exp \left( -\frac{I^{2/\delta}}{R T_{\mathrm{int}}} \right).
\end{split}
\end{equation}
Noting that
\begin{equation}
\frac{\gamma_{k+1}^{(m)}}{\gamma_{k}^{(m)}}
  = \frac{\Gamma(m + k + 2)}{\Gamma(m + k + 1)} \cdot
    \frac{\Gamma(k + 1)}{\Gamma(k + 2)}
  = \frac{m + k + 1}{k + 1},
\end{equation}
one finally obtains a simple expression:
\begin{equation}
\frac{\partial}{\partial \eta} \psi_{2,k,T_{\mathrm{int}}}
  \left( \frac{I^{2/\delta}}{R T_{\mathrm{int}}} \right)
= -(m + k + 1) \psi_{2,k+1,T_{\mathrm{int}}}
  \left( \frac{I^{2/\delta}}{R T_{\mathrm{int}}} \right).
\end{equation}
Thus the derivative of the basis function \eqref{eq:bas_fun} is
\begin{equation} \label{eq:bas_diff}
\begin{split}
\frac{\partial}{\partial \eta}
  \psi_{\alpha,k,T_{\mathrm{tr}},T_{\mathrm{int}}} &=
- (m+k+1) \psi_{\alpha,k+1,T_{\mathrm{tr}}, T_{\mathrm{int}}} \\
& \qquad + \sum_{d=1}^3 \left[
  \frac{\partial u_d}{\partial \eta}
    \psi_{\alpha+e_d,k,T_{\mathrm{tr}}, T_{\mathrm{int}}} +
  \frac{1}{2} \frac{\partial (R T_{\mathrm{tr}})}{\partial \eta}
    \psi_{\alpha+2e_d,k,T_{\mathrm{tr}}, T_{\mathrm{int}}}
\right].
\end{split}
\end{equation}
Here $\psi_{\alpha,k,T_{\mathrm{tr}},T_{\mathrm{int}}}$ stands for
\begin{equation}
\psi_{\alpha,k,T_{\mathrm{tr}},T_{\mathrm{int}}} \left(
  \frac{\bxi - \bu}{\sqrt{R T_{\mathrm{tr}}}},
  \frac{I^{2/\delta}}{R T_{\mathrm{int}}}
\right).
\end{equation}
The parameters are omitted for conciseness.

Now we expand the left hand side of \eqref{eq:ES-BGK} into series.
Using \eqref{eq:bas_diff}, one immediately has
\begin{equation} \label{eq:df_dt}
\begin{split}
\frac{\partial f}{\partial t} &=
  \sum_{\alpha \in \bbN^3} \sum_{k \in \bbN} \left[
    \frac{\partial f_{\alpha,k}}{\partial t}
      \psi_{\alpha,k,T_{\mathrm{tr}},T_{\mathrm{int}}} +
    f_{\alpha,k} \frac{\partial}{\partial t}
      \psi_{\alpha,k,T_{\mathrm{tr}},T_{\mathrm{int}}}
  \right] \\
&= \sum_{\alpha \in \bbN^3} \sum_{k \in \bbN} \Bigg[
    \frac{\partial f_{\alpha,k}}{\partial t} +
    \sum_{d=1}^3 \frac{\partial u_d}{\partial t} f_{\alpha+e_d, k} \\
& \qquad \qquad \qquad +\frac{1}{2}
    \frac{\partial (R T_{\mathrm{tr}})}{\partial t}
      \sum_{d=1}^3 f_{\alpha - 2e_d, k} -
    (m + k) \frac{\partial (R T_{\mathrm{int}})}{\partial t}
      f_{\alpha, k-1}
  \Bigg] \psi_{\alpha,k,T_{\mathrm{tr}},T_{\mathrm{int}}}.
\end{split}
\end{equation}
For the convection term, we have
\begin{equation}
\begin{split}
\bxi \cdot \nabla_{\bx} f &=
  \sum_{j=1}^3 \xi_j \frac{\partial f}{\partial x_j} =
  \sum_{j=1}^3 \xi_j \sum_{\alpha \in \bbN^3} \sum_{k \in \bbN} \Bigg[
    \frac{\partial f_{\alpha,k}}{\partial x_j} +
    \sum_{d=1}^3 \frac{\partial u_d}{\partial x_j} f_{\alpha+e_d, k}\\
& \qquad \qquad \qquad +\frac{1}{2}
    \frac{\partial (R T_{\mathrm{tr}})}{\partial x_j}
      \sum_{d=1}^3 f_{\alpha - 2e_d, k} -
    (m + k) \frac{\partial (R T_{\mathrm{int}})}{\partial x_j}
      f_{\alpha, k-1}
  \Bigg] \psi_{\alpha,k,T_{\mathrm{tr}},T_{\mathrm{int}}}.
\end{split}
\end{equation}
Now we use \eqref{eq:times_xi} and get
\begin{equation} \label{eq:xi_df_dx}
\begin{split}
\bxi \cdot \nabla_{\bx} f &=
\sum_{\alpha \in \bbN^3} \sum_{k \in \bbN} \sum_{j=1}^3 \bigg[ \left(
  R T_{\mathrm{tr}} \frac{\partial f_{\alpha-e_j,k}}{\partial x_j}
  + u_j \frac{\partial f_{\alpha,k}}{\partial x_j}
  + (\alpha_j + 1) \frac{\partial f_{\alpha+e_j,k}}{\partial x_j}
\right) \\
& \quad + \sum_{d=1}^3 \frac{\partial u_d}{\partial x_j}
  \left( R T_{\mathrm{tr}} f_{\alpha-e_d-e_j,k} + u_j f_{\alpha-e_d,k}
    + (\alpha_j + 1) f_{\alpha-e_d+e_j,k} \right) \\
& \quad + \frac{1}{2}
  \frac{\partial (R T_{\mathrm{tr}})}{\partial x_j} \sum_{d=1}^3 \left(
    R T_{\mathrm{tr}} f_{\alpha-2e_d-e_j,k}
    + u_j f_{\alpha-2e_d,k}
    + (\alpha_j + 1) f_{\alpha-2e_d+e_j,k}
  \right) \\
& \quad -(m+k)\frac{\partial (R T_{\mathrm{int}})}{\partial x_j}
  \left(
    R T_{\mathrm{tr}} f_{\alpha-e_j,k-1}
    + u_j f_{\alpha,k-1}
    + (\alpha_j + 1) f_{\alpha+e_j,k-1}
  \right) \bigg] \psi_{\alpha,k,T_{\mathrm{tr}},T_{\mathrm{int}}}.
\end{split}
\end{equation}
Collecting \eqref{eq:df_dt}\eqref{eq:xi_df_dx} and
\eqref{eq:G_expand}, the moment system \eqref{eq:mnt_eqs} follows
naturally.

\section{Expansion of the generalized Gaussian} \label{sec:G}
This section is devoted to the calculation of $G_{\alpha,0}$, which is
defined in \eqref{eq:G_expand}. In this appendix, $G$ is considered as
a function of $\bxi$ and $I$, where the parameters $t$ and $\bx$ are
omitted. It can be deduced from the orthogonality of Hermite and
Laguerre polynomials that
\begin{equation} \label{eq:G_alpha_0}
G_{\alpha,0} = C_{\alpha, T_{\mathrm{tr}}} \int_{\bbR^3 \times \bbR^+}
  p_{\alpha, T_{\mathrm{tr}}}(\bv)
  G \left(\bu + \sqrt{R T_{\mathrm{tr}}} \bv, I \right) \dd \bv \dd I,
\end{equation}
where $p_{\alpha, T_{\mathrm{tr}}}$ is a polynomial defined as
\begin{equation}
p_{\alpha, T_{\mathrm{tr}}}(\bv) =
  \psi_{1,\alpha,T_{\mathrm{tr}}}(\bv) \exp \left(
    -\frac{|\bv|^2}{2}
  \right) =
\left( \sqrt{2 \pi} \right)^{-3}
  (R T_{\mathrm{tr}})^{-\frac{|\alpha| + 3}{2}}
\prod_{d=1}^3 \He_{\alpha_d}(v_d),
\end{equation}
and $C_{\alpha, T_{\mathrm{tr}}}$ is a constant dependent on $\alpha$
and $T_{\mathrm{tr}}$:
\begin{equation}
C_{\alpha, T_{\mathrm{tr}}} =
  \frac{(2 \pi)^{-\frac{3}{2}} (R T_{\mathrm{tr}})^{3 + |\alpha|}}
    {\alpha_1! \alpha_2! \alpha_3!}.
\end{equation}
For $i \in \{1,2,3\}$, if $\alpha_i > 0$, the recursion relation of
Hermite polynomials shows that 
\begin{equation}
p_{\alpha, T_{\mathrm{tr}}}(\bv) =
  (R T_{\mathrm{tr}})^{-\frac{1}{2}} v_i
    p_{\alpha - e_i, T_{\mathrm{tr}}}(\bv) -
  (R T_{\mathrm{tr}})^{-1} (\alpha_i - 1)
    p_{\alpha - 2 e_i, T_{\mathrm{tr}}}(\bv).
\end{equation}
Noting that
\begin{equation} \label{eq:C_alpha}
C_{\alpha, T_{\mathrm{tr}}}
= \frac{(R T_{\mathrm{tr}})^2}{\alpha_i (\alpha_k - \delta_{ik})}
    C_{\alpha - e_i - e_k, T_{\mathrm{tr}}}
= \frac{(R T_{\mathrm{tr}})^2}{\alpha_i (\alpha_i - 1)}
    C_{\alpha - 2e_i, T_{\mathrm{tr}}},
\end{equation}
one directly obtains from \eqref{eq:G_alpha_0} that
\begin{equation}
G_{\alpha,0} = C_{\alpha, T_{\mathrm{tr}}}
  (R T_{\mathrm{tr}})^{-\frac{1}{2}} \int_{\bbR^3 \times \bbR^+}
    v_i p_{\alpha - e_i, T_{\mathrm{tr}}}(\bv)
    G \left(\bu + \sqrt{R T_{\mathrm{tr}}} \bv, I \right)
  \dd \bv \dd I
- \frac{R T_{\mathrm{tr}}}{\alpha_i} G_{\alpha - 2e_i, 0}.
\end{equation}
Now the expression of $G$ \eqref{eq:G} is put into the above equation.
After integrating with respect to $I$, one has
\begin{equation} \label{eq:G_alpha_0_split}
G_{\alpha, 0} =
  \frac{C_{\alpha, T_{\mathrm{tr}}} \rho (R T_{\mathrm{tr}})^{-\frac{1}{2}}}
    {\sqrt{\mathrm{det} (2 \pi \mathcal{T})}}
  \int_{\bbR^3} v_i p_{\alpha - e_i, T_{\mathrm{tr}}}(\bv) \exp\left(
    -\frac{R T_{\mathrm{tr}}}{2} \bv^T \mathcal{T}^{-1} \bv
  \right) \dd \bv
- \frac{R T_{\mathrm{tr}}}{\alpha_i} G_{\alpha - 2e_i, 0}.
\end{equation}
The matrix $\mathcal{T}$ is required to be positive definite, since
the density of the fluid should be finite. Thus, there exists a matrix
$\mathcal{R} = (r_{ij})$ such that $\mathcal{T} = (R T_{\mathrm{tr}})
\mathcal{R} \mathcal{R}^T$. Making the transformation $\bw =
\mathcal{R}^{-1} \bv$, and noting that $\mathrm{det}(\mathcal{T}) =
(R T_{\mathrm{tr}})^3 [\mathrm{det} (\mathcal{R})]^2$, \eqref
{eq:G_alpha_0_split} becomes
\begin{equation} \label{eq:rotated}
G_{\alpha,0} =
  \frac{C_{\alpha, T_{\mathrm{tr}}} \rho}
    {(\sqrt{2 \pi})^3 (R T_{\mathrm{tr}})^2}
  \sum_{j=1}^3 r_{ij} \int_{\bbR^3}
    w_j p_{\alpha - e_i, T_{\mathrm{tr}}} (\mathcal{R} \bw)
    \exp (-|\bw|^2 / 2)
  \dd \bw - \frac{R T_{\mathrm{tr}}}{\alpha_i} G_{\alpha - 2e_i, 0}.
\end{equation}
The following relation is a direct result of the differential relation
of Hermite polynomials:
\begin{equation}
\frac{\partial}{\partial w_j}
  p_{\alpha, T_{\mathrm{tr}}}(\mathcal{R} \bw)
= (R T_{\mathrm{tr}})^{-\frac{1}{2}}
  \sum_{k=1}^3 \alpha_k r_{kj}
    p_{\alpha - e_k, T_{\mathrm{tr}}}(\mathcal{R} \bw).
\end{equation}
Thus it can be obtained by integrating by parts that
\begin{equation} \label{eq:ibp}
\begin{split}
& \int_{\bbR^3} w_j p_{\alpha - e_i, T_{\mathrm{tr}}} (\mathcal{R}\bw)
  \exp ( -|\bw|^2 / 2 ) \dd \bw \\
={} & (R T_{\mathrm{tr}})^{-\frac{1}{2}} \sum_{k=1}^3
  (\alpha_k - \delta_{ik}) r_{kj}
  \int_{\bbR^3} p_{\alpha-e_i-e_k, T_{\mathrm{tr}}}(\mathcal{R} \bw)
    \exp (-|\bw|^2 / 2) \dd \bw.
\end{split}
\end{equation}
Now we substitute \eqref{eq:ibp} into \eqref{eq:rotated}, and apply
the transformation $\bv = \mathcal{R} \bw$. The result is
\begin{equation}
\begin{split}
G_{\alpha,0} &= \frac{\rho}{\sqrt{\mathrm{det} (2 \pi \mathcal{T})}}
  \sum_{j=1}^3 r_{ij} \sum_{k=1}^3 r_{kj} \cdot
  \frac{C_{\alpha, T_{\mathrm{tr}}} (\alpha_k - \delta_{ik})}
    {R T_{\mathrm{tr}}} \times \\
& \qquad \qquad \qquad
  \int_{\bbR^3} p_{\alpha-e_i-e_k, T_{\mathrm{tr}}}(\bv) \exp \left(
    -\frac{R T_{\mathrm{tr}}}{2} \bv^T \mathcal{T}^{-1} \bv
  \right) \dd \bv
- \frac{R T_{\mathrm{tr}}}{\alpha_i} G_{\alpha - 2e_i, 0}.
\end{split}
\end{equation}
Using \eqref{eq:C_alpha}, it is not difficult to find
\begin{equation}
G_{\alpha,0} = \frac{R T_{\mathrm{tr}}}{\alpha_i} \left(
  \sum_{j=1}^3 r_{ij} \sum_{k=1}^3 r_{kj} G_{\alpha-e_i-e_k}
  - G_{\alpha - 2e_i, 0}
\right).
\end{equation}
Recalling $\mathcal{T} = (R T_{\mathrm{tr}}) \mathcal{R}
\mathcal{R}^T$, the above equation can be written as
\begin{equation} \label{eq:G_alpha_0_recur}
G_{\alpha,0} = \frac{1}{\alpha_i} \sum_{k=1}^3
  (\lambda_{ik} - R T_{\mathrm{tr}} \delta_{ik})
  G_{\alpha - e_i - e_k,0},
\end{equation}
where $\lambda_{ik}$ is the $(i,k)$-element of $\mathcal{T}$. The
final result \eqref{eq:G_recur} is then obtained by substituting the
detailed expression of $\mathcal{T}$ into \eqref{eq:G_alpha_0_recur}.

\bibliographystyle{plain}
\bibliography{../article}

\begin{thebibliography}{10}

\bibitem{Abramowitz}
M.~Abramowitz and I.~A. Stegun.
\newblock {\em Handbook of Mathematical Functions with Formulas, Graphs, and
  Mathematical Tables}.
\newblock Dover, New York, 1964.

\bibitem{Alsmeyer}
H.~Alsmeyer.
\newblock Density profiles in argon and nitrogen shock waves measured by the
  absorption of an electron beam.
\newblock {\em J. Fluid. Mech.}, 74(3):497--513, 1976.

\bibitem{Bourgat}
P.~Andries, J.~F. Bourgat, P.~L. Tallec, and B.~Perthame.
\newblock Numerical comparison between the {Boltzmann} and {ES}-{BGK} models
  for rarefied gases.
\newblock {\em Comput. Methods Appl. Mech. Engrg.}, 191(31):3369--3390, 2002.

\bibitem{Andries}
P.~Andries, P.~L. Tallec, J.~P. Perlat, and B.~Perthame.
\newblock The {G}aussian-{BGK} model of {B}oltzmann equation with small
  {P}randtl number.
\newblock {\em Eur. J. Mech. B - Fluids}, 19(6):813--830, 2000.

\bibitem{Weiss}
J.~D. Au, M.~Torrilhon, and W.~Weiss.
\newblock The shock tube study in extended thermodynamics.
\newblock {\em Phys. Fluids}, 13(8):2423--2432, 2001.

\bibitem{Brull}
S.~Brull and J.~Schneider.
\newblock On the ellipsoidal statistical model for polyatomic gases.
\newblock {\em Continuum Mech. Thermodyn.}, 20(8):489--508, 2009.

\bibitem{NRxx}
Z.~Cai and R.~Li.
\newblock Numerical regularized moment method of arbitrary order for
  {B}oltzmann-{BGK} equation.
\newblock {\em SIAM J. Sci. Comput.}, 32(5):2875--2907, 2010.

\bibitem{Cai}
Z.~Cai, R.~Li, and Y.~Wang.
\newblock An efficient {\NRxx} method for {B}oltzmann-{BGK} equation.
\newblock {\em J. Sci. Comput.}, 50(1):103--119, 2012.

\bibitem{NRxx_new}
Z.~Cai, R.~Li, and Y.~Wang.
\newblock Numerical regularized moment method for high {M}ach number flow.
\newblock {\em Commun. Comput. Phys.}, 11(5):1415--1438, 2012.

\bibitem{Dubroca}
B.~Dubroca and L.~Mieussens.
\newblock A conservative and entropic discrete-velocity model for rarefied
  polyatomic gases.
\newblock In {\em CEMRACS 1999 (Orsay)}, volume~10 of {\em ESAIM Proc.}, pages
  127--139, Paris, 1999. Soc. Math. Appl. Indust.

\bibitem{Grad}
H.~Grad.
\newblock On the kinetic theory of rarefied gases.
\newblock {\em Comm. Pure Appl. Math.}, 2(4):331--407, 1949.

\bibitem{Grad1952}
H.~Grad.
\newblock The profile of a steady plane shock wave.
\newblock {\em Comm. Pure Appl. Math.}, 5(3):257--300, 1952.

\bibitem{Larina}
I.~N. Larina and V.~A. Rykov.
\newblock Kinetic model of the {B}oltzmann equation for a diatomic gas with
  rotational degrees of freedom.
\newblock {\em Comput. Math. Math. Phys.}, 50(12):2118--2130, 2010.

\bibitem{Mallinger}
F.~Mallinger.
\newblock Generalization of the {G}rad theory to polyatomic gases.
\newblock Rapport de recherche 3581, INRIA Rocquencourt, 1998.

\bibitem{McCormack}
F.~J. McCormack.
\newblock Kinetic equations for polyatomic gases: The 17-moment approximation.
\newblock {\em Phys. Fluids}, 11(12):2533--2543, 1968.

\bibitem{Mieussens}
L.~Mieussens.
\newblock Discrete velocity model and implicit scheme for the {BGK} equation of
  rarefied gas dynamics.
\newblock {\em Math. Models Methods Appl. Sci.}, 10(8):1121--1149, 2000.

\bibitem{Reitebuch}
I.~M{\"u}ller, D.~Reitebuch, and W.~Weiss.
\newblock Extended thermodynamics -- consistent in order of magnitude.
\newblock {\em Continuum Mech. Thermodyn.}, 15(2):113--146, 2002.

\bibitem{Rykov}
V.~A. Rykov.
\newblock A model kinetic equation for a gas with rotational degrees of
  freedom.
\newblock {\em Fluid Dyn.}, 10(6):959--966, 1975.

\bibitem{Struchtrup2004}
H.~Struchtrup.
\newblock Stable transport equations for rarefied gases at high orders in the
  {K}nudsen number.
\newblock {\em Phys. Fluids}, 16(11):3921--3934, 2004.

\bibitem{Struchtrup}
H.~Struchtrup.
\newblock {\em Macroscopic Transport Equations for Rarefied Gas Flows:
  Approximation Methods in Kinetic Theory}.
\newblock Springer, 2005.

\bibitem{Struchtrup2003}
H.~Struchtrup and M.~Torrilhon.
\newblock Regularization of {G}rad's 13 moment equations: Derivation and linear
  analysis.
\newblock {\em Phys. Fluids}, 15(9):2668--2680, 2003.

\bibitem{Torrilhon2000}
M.~Torrilhon.
\newblock Characteristic waves and dissipation in the 13-moment-case.
\newblock {\em Continuum Mech. Thermodyn.}, 12(5):289--301, 2000.

\bibitem{Au}
M.~Torrilhon, J.~Au, D.~Reitebuch, and W.~Weiss.
\newblock The {R}iemann-problem in extended thermodynamics.
\newblock In H.~Freistu{\"u}hler and G.~Warnecke, editors, {\em Hyperbolic
  Problems: Theory, Numerics, Applications, Vols {I} and {II}}, volume 140 of
  {\em International series of numerical mathematics}, pages 79--88.
  Birkh{\"a}user, 2001.

\bibitem{Tang}
K.~Xu and L.~Tang.
\newblock Nonequilibrium {B}hatnagar-{G}ross-{K}rook model for nitrogen shock
  structure.
\newblock {\em Phys. Fluids}, 16(10):3824--3827, 2004.

\end{thebibliography}
\end{document}